\documentclass[11pt,a4paper]{article}
\usepackage[T1]{fontenc}
\usepackage[utf8]{inputenc}
\usepackage[margin=21mm,headheight=14pt,footskip=11mm]{geometry}
\usepackage{lmodern}
\usepackage{mathpazo}
\usepackage[dvipsnames]{xcolor}
\definecolor{ink}{RGB}{38,48,58}
\definecolor{muted}{RGB}{85,92,99}
\usepackage{amsmath,amssymb,graphicx,float,makecell,multirow,booktabs,tabularx}
\usepackage{enumitem,changepage,microtype,algorithm,algpseudocode}
\usepackage{placeins}
\usepackage{tikz}
\usetikzlibrary{arrows,positioning,calc,decorations.pathreplacing}
\usepackage[numbers,sort&compress]{natbib}
\setcitestyle{citesep={,}}

\usepackage{caption}
\usepackage{titlesec}
\titleformat{\section}{\large\bfseries\color{ink}}{\thesection.}{0.6em}{}
\titleformat{\subsection}{\normalsize\bfseries\color{ink}}{\thesubsection.}{0.6em}{}
\titleformat{\subsubsection}{\normalsize\itshape}{\thesubsubsection.}{0.6em}{}
\titlespacing*{\section}{0pt}{2.2ex plus .5ex minus .2ex}{1ex}
\titlespacing*{\subsection}{0pt}{1.8ex plus .4ex minus .2ex}{.7ex}
\titlespacing*{\subsubsection}{0pt}{1.4ex plus .3ex minus .2ex}{.5ex}
\titlespacing*{\paragraph}{0pt}{1.2ex plus .2ex}{.7em}
\usepackage[hidelinks,unicode]{hyperref}
\hypersetup{pdftitle={Enhancing Photogrammetric Digital Surface Models with Pretrained Diffusion Models and Multimodal Conditioning},pdfauthor={Antoine Lorentz, Stéphane May, Valentine Bellet, Dawa Derksen and Bastien Nespoulous}}
\newcolumntype{C}{>{\centering\arraybackslash}X}
\newcolumntype{L}{>{\raggedright\arraybackslash}X}
\newcolumntype{R}{>{\raggedleft\arraybackslash}X}
\newlength{\extralength}
\newlength{\fulllength}
\newcommand{\backmatteritem}[2]{\par\smallskip\noindent\textbf{#1:} #2\par}
\newcommand{\backmattersection}[2]{\section*{#1}#2\par}

\setlist{itemsep=2pt,parsep=0pt,topsep=4pt,leftmargin=1.5em}
\begin{document}
\begin{center}
{\fontsize{17}{21}\selectfont\bfseries\color{ink}
Enhancing Photogrammetric Digital Surface Models with\\
Pretrained Diffusion Models and Multimodal Conditioning\par}
\vspace{9pt}
\href{https://orcid.org/0009-0001-7042-7558}{Antoine Lorentz}\textsuperscript{1,*}, \href{https://orcid.org/0000-0002-4845-7232}{Stéphane May}\textsuperscript{2}, \href{https://orcid.org/0000-0003-3679-3631}{Valentine Bellet}\textsuperscript{2},\\[2pt]
Dawa Derksen\textsuperscript{2} and Bastien Nespoulous\textsuperscript{1}
\end{center}
\begin{center}
\begin{minipage}{.96\textwidth}
\fontsize{9}{11}\selectfont\color{muted}\raggedright
\textsuperscript{1} Thales, 92098 Paris, France; bastien.nespoulous@thalesgroup.com\\[2pt]
\textsuperscript{2} Centre National d'Études Spatiales (CNES), 75039 Paris, France; stephane.may@cnes.fr (S.M.); valentine.bellet@cnes.fr (V.B.); dawa.derksen@cnes.fr (D.D.)\\[2pt]
* Correspondence: antoine.lorentz@thalesgroup.com
\end{minipage}
\end{center}
\begin{center}\fontsize{9}{11}\selectfont\color{muted}
Published in \emph{Remote Sensing} \textbf{2026}, 18, 3303.
Version of record: \url{https://doi.org/10.3390/rs18193303}.\\
Author-formatted version. © 2026 the authors. \href{https://creativecommons.org/licenses/by/4.0/}{CC BY 4.0}.
\end{center}
\noindent{\color{ink!30}\rule{\textwidth}{.4pt}}
\section*{Highlights}
\noindent
\begin{minipage}[t]{.48\textwidth}
\small\raggedright
\textbf{What are the main findings?}
\begin{itemize}
    \item Patch-wise normalization enables the stable adaptation of a pretrained diffusion model to 3D elevation maps with widely varying mean elevations and local elevation variability.
    \item Conditioning the model on a photogrammetric Digital Surface Model and Pléiades-HR optical imagery improves elevation refinement, reducing Dense Urban RMSE from 6.00 to 3.45\,m across eight French test cities and from 4.16 to 2.77\,m in the geographically held-out city of Bordeaux.
\end{itemize}
\end{minipage}\hfill
\begin{minipage}[t]{.48\textwidth}
\small\raggedright
\textbf{What are the implications of the main findings?}
\begin{itemize}
    \item Visual representations learned from large-scale natural-image pretraining can be transferred to the structurally different domain of metric elevation-map generation and combined with multiple geospatial modalities.
    \item Patch-wise normalization provides a practical approach for adapting diffusion and flow-matching models to physical-valued raster data whose offsets and dynamic ranges vary substantially between samples, while restoring predictions in their original physical~units.
\end{itemize}
\end{minipage}
\section*{Abstract}
Large-scale Digital Surface Models (DSMs) can be produced cost-effectively from satellite images via stereo-photogrammetry. However, the resulting 3D maps are often contaminated by noise, outliers, and voids. On the other hand, aerial LiDAR provides high-accuracy elevation measurements at a substantially higher cost. In this work, we study diffusion models conditioned both on photogrammetric DSMs and Pléiades imagery to refine vertically co-registered DSMs. We introduce a modified Stable Diffusion 3 architecture with a pruned text stream and a patch-wise normalization strategy, enabling stable training on LiDAR data and transfer from natural images to elevation maps. Experiments in French cities demonstrate that multimodal conditioning improves elevation accuracy, reducing Dense Urban RMSE from 6.00 to 3.45\,m in the in-context cities and from 4.16 to 2.77\,m in the held-out city of Bordeaux.

\smallskip
{\small\noindent\textbf{Keywords:} DSM enhancement; diffusion models; flow matching; multimodal conditioning; Stable Diffusion 3; geospatial data\par}
\section{Introduction}

State-of-the-art stereo-photogrammetry pipelines (e.g., CARS~\cite{youssefi_cars_2020}, S2P~\cite{de_franchis_automatic_2014,de_franchis_stereo-rectification_2014,de_franchis_automatic_2014-1}, MicMac~\cite{rupnik_micmac_2017}, and NASA ASP~\cite{beyer_ames_2018}) offer a cost-effective solution for large-scale Digital Surface Model (DSM) production from optical satellite imagery. However, these DSMs typically suffer from high noise levels, incomplete coverage due to occlusions and low-texture regions, and errors stemming from matching ambiguities. On~the other hand, aerial LiDAR provides high-accuracy measurements, but its substantially higher cost and limited availability make it less suitable for large-scale, up-to-date mapping, especially in remote areas. This motivates methods that refine photogrammetric DSMs toward LiDAR-quality elevation.

Recent advances in deep generative models, particularly diffusion models, have revolutionized image synthesis. Flow matching~\cite{lipman2023flow} and rectified flows~\cite{liu2022flow} provide efficient frameworks for learning probability transport between noise and data distributions. When scaled to large transformer architectures~\cite{peebles_scalable_2023}, these models demonstrate remarkable capabilities in generating high-fidelity images~\cite{esser2024scaling}. Latent diffusion models~\cite{rombach2022high} further improve computational efficiency by operating in a compressed latent space. Critically, ControlNet~\cite{zhang_adding_2023} introduced a paradigm for conditioning these powerful generative models on pixel-wise guidance, enabling spatially localized control over the generated content while preserving the rich visual priors learned from large-scale pretraining.

The application of diffusion models to terrain and elevation data has gained significant momentum. Diff-DEM~\cite{shih-huang_lo_diff-dem_2024} demonstrated that diffusion models excel at filling voids in Digital Elevation Models (DEMs), a task also addressed using deterministic, non-generative edge-enhancing diffusion~\cite{panangian_dfilled_2025}, while MESA~\cite{borne--pons_mesa_2025} proposed text-driven terrain generation using a modified pretrained diffusion model. These works establish diffusion models as powerful tools for terrain generation, but they primarily focus on void filling or text-driven generation rather than the comprehensive refinement of noisy photogrammetric~data.

Our study also relates to depth completion, multimodal fusion, and GAN-based elevation restoration. Recent advances include Marigold-DC~\cite{viola2025marigold}, which achieves zero-shot depth completion by dynamically incorporating sparse measurements during inference. GAN-based DEM methods have also addressed void filling using contextual attention or terrain-texture priors, including a Wasserstein GAN~\cite{gavriil2019void}, a terrain-texture model~\cite{qiu2019void}, a~context-attention model~\cite{zhang2020dem}, and a multiattention model~\cite{zhou2022voids}. These methods are relevant to our secondary void-recovery capability, but their inputs and evaluation protocols target missing DEM regions rather than systematic correction of a dense stereo DSM. In remote sensing and geospatial applications, SatelliteMaker~\cite{yu_diffusion-based_2025} employs ControlNet with DEMs to guide satellite image reconstruction, highlighting the value of elevation conditioning. Methods using texture transfer from high-resolution images have shown promise for DSM super-resolution~\cite{ye_dem_2024}, while Wang et al. demonstrated improvements in super-resolution through terrain trends and residual decomposition~\cite{wang_dem_2024}. The DSM-to-LoD2~\cite{bittner_dsm--lod2_2018} approach used GANs to refine stereo DSMs against structured CityGML references.

These approaches are not directly interchangeable baselines for the present study. Void-completion methods primarily target missing regions in masked elevation inputs; super-resolution methods primarily change spatial resolution; Marigold-DC uses sparse depth. DSM-to-LoD2 refines dense stereo DSMs into LoD2-like raster elevations using CityGML-derived references. Table~\ref{tab:task_matrix} summarizes these task boundaries. We use the calibrated stereo DSM input as the task baseline and evaluate our conditional refinement model within the same paired protocol, under two explicit assumptions on the input DSM:

\begin{description}
    \item[(A1) Vertical co-registration] The DSM has been aligned to the LiDAR vertical datum at the acquisition level, removing any acquisition-wide vertical offset (Section~\ref{sec:data}).
    \item[(A2) Surface compatibility] The DSM and LiDAR are assumed to represent the same underlying physical surface. This is a statement about the surface, not about acquisition dates: two close-in-time acquisitions can still violate A2 (e.g.,\ a building is demolished or a canopy becomes fully leafed out between passes), while two distant-in-time acquisitions can satisfy it if the surface is stable. We do not verify A2 directly; we only apply the coarse, DSM-error-based curation described in Section~\ref{sec:data}.
\end{description}

The estimand is therefore the correction of local errors in a DSM satisfying A1--A2, not general DSM-to-LiDAR reconstruction from an arbitrary, uncalibrated, or surface-inconsistent input. Section~\ref{sec:limitations} returns to cases where A2 is plausibly violated, in particular seasonal vegetation change.
\label{sec:task_assumptions}

\begin{table}[!htbp]
\caption{Task comparison used to define the internal baseline. A checkmark denotes a primary or optional input/task association; a star marks a secondary/optional capability; a cross denotes a different target or input setting.}
\label{tab:task_matrix}
\footnotesize
\setlength{\tabcolsep}{3pt}

\begin{adjustwidth}{-\extralength}{0cm}
\begin{tabularx}{\fulllength}{lCCCCCC}
\toprule
\textbf{Method} & \makecell{\textbf{Dense DSM}\\\textbf{Correction}} & \makecell{\textbf{Void}\\\textbf{Filling}} & \makecell{\textbf{Sparse-Depth}\\\textbf{Completion}} & \makecell{\textbf{Resolution}\\\textbf{Change}} & \makecell{\textbf{LoD2-like}\\\textbf{Target}} & \makecell{\textbf{RGB}\\\textbf{Conditioning}} \\
\midrule
Diff-DEM & $\times$ & $\checkmark$ & $\times$ & $\times$ & $\times$ & $\times$ \\
Dfilled & $\times$ & $\checkmark$ & $\times$ & $\times$ & $\times$ & $\checkmark$ \\
\makecell[l]{GAN void filling \cite{gavriil2019void,qiu2019void,zhang2020dem,zhou2022voids}} & $\times$ & $\checkmark$ & $\times$ & $\times$ & $\times$ & $\times$ \\
Marigold-DC & $\times$ & $\times$ & $\checkmark$ & $\times$ & $\times$ & $\checkmark$ \\
DSM super-resolution & $\times$ & $\times$ & $\times$ & $\checkmark$ & $\times$ & $\checkmark$ {*} \\
DSM-to-LoD2 & $\checkmark$ & $\times$ & $\times$ & $\times$ & $\checkmark$ & $\times$ \\
This work & $\checkmark$ & $\checkmark$ {*} & $\times$ & $\times$ & $\times$ & $\checkmark$ \\
\bottomrule
\end{tabularx}
\end{adjustwidth}
\end{table}

In this study, we address these challenges by adapting Stable Diffusion 3 (SD3)~\cite{esser2024scaling} for DSM enhancement through multimodal conditioning on photogrammetric DSMs and Pléiades-HR optical imagery. Our key contributions are three-fold. First, we introduce a patch-wise normalization strategy that stabilizes training on locally low-variance elevation data. Second, we develop an image-only architecture by pruning the text stream from SD3's multimodal transformer, reducing its transformer parameter count from approximately 2B to 1B while retaining pretrained visual representations. Third, we use separate ControlNets to condition on DSMs and optical imagery, and evaluate the resulting DSM-only and DSM~+~RGB configurations within the same paired protocol.

Our experiments use a dataset combining LiDAR-HD {(\url{https://cartes.gouv.fr/telechargement/IGNF\_NUAGES-DE-POINTS-LIDAR-HD}, accessed on 1 December 2025)} from 20 French cities with Pléiades-HR imagery and CARS stereo DSMs covering 9 of these cities. Bordeaux is held out as a geographically isolated, in-country test city. We assess adaptation of the pretrained SD3 backbone in this French urban setting; transfer to other countries, sensors, processing chains, and terrain types remains for future evaluation.

\FloatBarrier
\section{Materials and Methods}

\subsection{Flow Matching Background}

Flow matching~\cite{lipman2023flow,liu2022flow} learns a continuous transformation between a noise distribution $p_0$ and a data distribution $p_1$ through an Ordinary Differential Equation (ODE):
\begin{equation}
\frac{d}{dt} x_t = v_t(x_t),
\end{equation}
where the time-dependent velocity field $v_t$ transports probability mass along a density path $p_t$ from $p_0$ to $p_1$.

In multisample flow matching~\cite{pooladian2023multisample}, the model is trained by sampling pairs $(x_0,x_1)$ from a joint distribution $q(x_0,x_1)$ whose marginals match the source and target distributions:
\begin{equation}
\int q(x_0,x_1)\,dx_1 = p_0(x_0),
\qquad
\int q(x_0,x_1)\,dx_0 = p_1(x_1).
\end{equation}

This formulation allows flexibility in defining the coupling between $x_0$ and $x_1$. Importantly, $x_0$ and $x_1$ do not need to be independent. Introducing dependence between them can reduce the variance in the training objective and produce straighter transport trajectories.

Given samples $(x_0,x_1) \sim q(x_0,x_1)$, a trajectory $x_t$ is constructed using a deterministic interpolation function $f$:
\begin{equation}
 x_t = f(x_0,x_1,t).
\end{equation}

For the commonly used rectified flow parameterization, the path is a straight line:
\begin{equation}
 x_t = (1-t)x_0 + t x_1.
\end{equation}

The corresponding conditional velocity field is
\begin{equation}
 v_t(x_t|x_1) = x_1 - x_0.
\end{equation}

The neural network $v_\theta$ is trained to match this velocity field through the Joint Conditional Flow Matching (JCFM) objective:
\begin{equation}
\mathcal{L}_{\mathrm{JCFM}}(\theta) =
\mathbb{E}_{t \sim \mathcal{U}[0,1],\,(x_0,x_1) \sim q}
\left\|
 v_\theta(x_t,t) - (x_1 - x_0)
\right\|^2.
\end{equation}

Once the velocity field is learned, new samples can be generated by integrating the ODE starting from $x_0 \sim p_0$:
\begin{equation} \label{eq:sampling}
 x_1 = x_0 + \int_0^1 v_t\,dt
\approx
 x_0 + \int_0^1 v_\theta\,dt.
\end{equation}

\subsection{Generating Elevation Maps}

Directly modeling transport from standard multivariate Gaussian noise to raw elevation patches $x_1$ causes instability during training. This instability arises because elevation variation within local terrain patches is typically small compared with the elevation range across the dataset, leading to poorly scaled training targets.
To mitigate this issue, we introduce a \textit{patch-wise noise scaling} scheme. We define two scalar random variables: a scale $s$ and an offset $u$. Instead of drawing the source noise $x_0$ independently from a standard normal distribution, we define a joint distribution $q(x_0, x_1)$, where the noise distribution is shifted and scaled to match the local characteristics of the target data:
\begin{equation}
    x_1 \sim p_1, \qquad x_0 \sim q(x_0|x_1) = \mathcal{N}(u \mathbf{1},\,s^2 I).
\end{equation}

By reparameterizing with standard Gaussian noise $\epsilon \sim \mathcal{N}(0,I)$, we can express the source noise as $x_0=s\epsilon+u$. Correspondingly, we define the normalized target data as $\hat{x}_1=(x_1-u)/s$, so that $x_1=s\hat{x}_1+u$.

The specific choice of $s$ and $u$ depends on the training phase:
\begin{itemize}
    \item \textbf{LiDAR-only backbone adaptation:} When learning the distribution using only LiDAR data, we define $u=\operatorname{mean}(x_1)$ and $s=\operatorname{std}(x_1)/0.25$. This normalizes the data to a scale similar to standard image datasets (Algorithm~\ref{alg:pretraining}).
    \item \textbf{Conditional modeling:} When modeling the target conditioned on a stereo DSM, we extract $u=\operatorname{mean}(x_{\mathrm{dsm}})$ and $s=\operatorname{std}(x_{\mathrm{dsm}})/0.25$ directly from the conditioning input (Algorithm~\ref{alg:conditional_training}). These statistics are computed over finite DSM pixels. At inference time, only the stereo DSM is required to derive them.
\end{itemize}
No epsilon, clipping, or zero-variance fallback is used; no instability attributable to patch-wise scaling was observed in the reported training runs. The minimum raw standard deviation in the scanned LiDAR support was $0.00473$\,m. The scan coverage and city distributions of $s$ and $u$ are reported in Appendix~\ref{app:data_support}.

\begin{algorithm}[!htbp]
\caption{LiDAR Backbone Pretraining}
\label{alg:pretraining}
\begin{algorithmic}[1]
\Require LiDAR dataset $\mathcal{D}$, scaling factor $\mathbb{E}[s^2]$

\Statex
\Repeat
    \State $x_1 \sim \mathcal{D}$

    \Statex
    \State $u \gets \operatorname{mean}(x_1)$
    \State $s \gets \operatorname{std}(x_1)/0.25$
    \State $\hat x_1 \gets (x_1-u)/s$

    \Statex
    \State $\epsilon \sim \mathcal{N}(0,I)$; \quad $t \sim \mathcal{U}(0,1)$;
    \State $\hat x_t \gets (1-t)\epsilon + t\hat x_1$

    \Statex
    \State $\mathcal{L} \gets \dfrac{s^2}{\mathbb{E}[s^2]}
           \big\|\hat v_\theta(\hat x_t,t,s,u) - (\hat x_1-\epsilon)\big\|^2$
    \State $\theta \gets \operatorname{AdamWUpdate}(\theta,\nabla_\theta\mathcal{L})$
\Until{convergence}
\end{algorithmic}
\end{algorithm}

\begin{algorithm}[!htbp]
\caption{DSM-Conditioned Training}
\label{alg:conditional_training}
\begin{algorithmic}[1]
\Require Multimodal dataset $\mathcal{D}$, scaling factor $\mathbb{E}[s^2]$

\Statex
\Repeat
    \State $(x_1, x_\mathrm{dsm})  \sim \mathcal{D}$

    \Statex
    \State $M \gets \operatorname{isfinite}(x_\mathrm{dsm})$
    \State $u \gets \operatorname{mean}(x_\mathrm{dsm}[M])$;
    \State $s \gets \operatorname{std}(x_\mathrm{dsm}[M])/0.25$
    \State $\hat x_1 \gets (x_1-u)/s$
    \State $\hat x_\mathrm{dsm}[M] \gets (x_\mathrm{dsm}[M]-u)/s$
    \State $\hat x_\mathrm{dsm}[\neg M] \gets -1$ \Comment{sentinel value for DSM voids}

    \Statex
    \State $\epsilon \sim \mathcal{N}(0,I)$; \quad $t \sim \mathcal{U}(0,1)$;
    \State $\hat x_t \gets (1-t)\epsilon + t\hat x_1$

    \Statex
    \State $\mathcal{L} \gets \dfrac{s^2}{\mathbb{E}[s^2]}
           \big\|\hat v_\theta(\hat x_t,t,s,u,\hat x_\mathrm{dsm}) - (\hat x_1-\epsilon)\big\|^2$
    \State $\theta \gets \operatorname{AdamWUpdate}(\theta,\nabla_\theta\mathcal{L})$
\Until{convergence}
\end{algorithmic}
\end{algorithm}

\subsubsection{Deriving the Loss Function}
Using the rectified flow formulation, the transport trajectory is a straight line: $x_t=(1-t)x_0+t x_1$. Substituting our scaled variables yields the following:
\begin{equation}
    x_t=(1-t)(s\epsilon+u)+t(s\hat{x}_1+u)=s\big[(1-t)\epsilon+t\hat{x}_1\big]+u=s\hat{x}_t+u,
\end{equation}
where $\hat{x}_t=(1-t)\epsilon+t\hat{x}_1$ represents the trajectory in the normalized space.

The target velocity field is the derivative of the path:
\begin{equation}
    v_t(x_t)=x_1-x_0=(s\hat{x}_1+u)-(s\epsilon+u)=s(\hat{x}_1-\epsilon).
\end{equation}

We reparameterize our neural network to predict the velocity in the normalized space, denoted as $\hat{v}_\theta(\hat{x}_t,t,s,u)\approx\hat{x}_1-\epsilon$. The corresponding unnormalized model velocity is therefore $v_\theta(x_t,t)=s\hat{v}_\theta(\hat{x}_t,t,s,u)$.

Substituting this into the standard Joint Conditional Flow Matching (JCFM) objective, we obtain the following:
\begin{equation}
\begin{aligned}
    \mathcal{L}_{\mathrm{JCFM}}(\theta)&=\mathbb{E}_{t,x_1,\epsilon}\left\|v_\theta(x_t,t)-(x_1-x_0)\right\|^2 \\
    &=\mathbb{E}_{t,x_1,\epsilon}\left\|s\hat{v}_\theta(\hat{x}_t,t,s,u)-s(\hat{x}_1-\epsilon)\right\|^2 \\
    &=\mathbb{E}_{t,x_1,\epsilon}\left[s^2\left\|\hat{v}_\theta(\hat{x}_t,t,s,u)-(\hat{x}_1-\epsilon)\right\|^2 \right].
\end{aligned}
\end{equation}

Finally, we normalize this objective by the positive global scalar $\mathbb{E}[s^2]$, estimated once before training. This normalization changes only the overall numerical scale of the objective; it does not change the relative physical weighting or its minimizer. It keeps the loss approximately on the scale used when training SD3 on natural images, allowing us to use a conventional fine-tuning learning rate rather than an unnecessarily small rate dictated only by elevation units. The final training objective is therefore as follows:
\begin{equation}
    \mathcal{L}(\theta)=\mathbb{E}_{t\sim\mathcal{U}[0,1],\,\epsilon\sim\mathcal{N}(0,I),\,x_1\sim p_1}\left[\frac{s^2}{\mathbb{E}[s^2]}\left\|\hat{v}_\theta(\hat{x}_t,t,s,u)-(\hat{x}_1-\epsilon)\right\|^2 \right].
\end{equation}

\subsubsection{Deriving the Sampling Procedure}
Once the model is trained, we generate new samples by integrating the learned velocity field along the ODE. Starting from the standard flow matching sampling equation,
\begin{equation}
    x_1=x_0+\int_0^1 v_\theta(x_t,t)\,dt,
\end{equation}
we substitute $x_0=s\epsilon+u$ and our parameterized velocity $v_\theta(x_t,t)=s\hat{v}_\theta(\hat{x}_t,t,s,u)$:
\begin{equation}
\begin{aligned}
    x_1&=(s\epsilon+u)+\int_0^1s\hat{v}_\theta(\hat{x}_t,t,s,u)\,dt \\
    &=s\left[\epsilon+\int_0^1\hat{v}_\theta(\hat{x}_t,t,s,u)\,dt\right]+u.
\end{aligned}
\end{equation}
The model generates the target data $x_1$ by first integrating standard Gaussian noise $\epsilon$ through the normalized probability flow $\hat{v}_\theta$, and subsequently, applying an explicit denormalization step using the scale $s$ and offset $u$ (Algorithm~\ref{alg:dsm_inference}).

\begin{algorithm}[!htbp]
\caption{DSM-Conditioned Inference}
\label{alg:dsm_inference}
\begin{algorithmic}[1]
\Require Trained velocity field $\hat v_\theta$, number of integration steps $K$
\Require DSM $x_\mathrm{dsm}$ with nonzero standard deviation; patches failing this requirement should be skipped.

\Statex
\State $M \gets \operatorname{isfinite}(x_\mathrm{dsm})$
\State $u \gets \operatorname{mean}(x_\mathrm{dsm}[M])$
\State $s \gets \operatorname{std}(x_\mathrm{dsm}[M])/0.25$
\State $\hat x_\mathrm{dsm}[M] \gets (x_\mathrm{dsm}[M]-u)/s$
\State $\hat x_\mathrm{dsm}[\neg M] \gets -1$ \Comment{sentinel value for DSM voids}

\Statex
\State $\hat x_0 \sim \mathcal{N}(0,I)$
\For{$k=0,\ldots,K-1$}
    \State $t_k \gets k/K$; \quad $\Delta t \gets 1/K$
    \State $\hat x_{t_{k+1}} \gets \hat x_{t_k} + \Delta t\, \hat v_\theta(\hat x_{t_k},t_k,s,u,\hat x_\mathrm{dsm})$
\EndFor

\Statex
\State $x_1 \gets s\,\hat x_{t_K} + u$
\State \Return $x_1$

\end{algorithmic}
\end{algorithm}

\FloatBarrier
\subsection{Model Architecture}
For all experiments, we use Stable Diffusion 3 Medium (SD3)~\cite{esser2024scaling}, a dual-stream diffusion transformer mixing text and image modalities. As SD3 is an image generator, we adapt elevation patches to an RGB-like format. During training, the normalized single-channel patch $\hat{x}_1$ is duplicated across three channels, allowing the Variational AutoEncoder (VAE) to process it. At inference, the three decoded channels are averaged before conversion to meters; this VAE reconstruction is part of the evaluated pipeline. We use two sinusoidal encoders followed by MultiLayer Perceptrons (MLPs) to embed the conditioning variables $s$ and $u$, and add the resulting embeddings to the timestep embedding.

Because textual annotations are neither available nor necessary for our map generation task, we transform the multimodal SD3 model into an image-only generator by removing the text processing components and converting joint text-image attention into pure image self-attention.

\subsubsection{Text-Stream Pruning}
To obtain an efficient, image-only backbone while retaining pretrained image weights, we prune the text stream of the Multimodal Diffusion Transformer (MM-DiT) as follows:
\begin{itemize}
    \item Replace joint cross-attention blocks that attend over text tokens with self-attention over image tokens.
    \item Remove projection matrices $Q, K, V$ that are specific to text tokens, along with the corresponding feed-forward networks (FFNs), LayerNorm modules and modulation layers that serve the text stream.
    \item Discard the external text encoders (CLIP-L~\cite{radford2021learning}, CLIP-G~\cite{radford2021learning}, T5-XXL~\cite{raffel2020exploring}) and the text pooled embedding.
\end{itemize}

This reduction lowers the transformer size from 2B to 1B parameters while still leveraging the pretrained weights of SD3 for the image stream. Pruning yields approximately $1.4\times$ higher inference throughput at batch size 32 in the synthetic 50-step denoising and VAE-decoding benchmark, both with and without ControlNets (Appendix~\ref{app:efficiency}, Table~\ref{tab:app_pruning_benchmark}). \mbox{Figures~\ref{fig:model_archi} and~\ref{fig:single_stream_block}} illustrate the resulting architecture and the structure of SingleStream-DiT blocks.

\begin{figure}[!htbp]
\centering
\def\figureheight{9cm}
\begin{minipage}[t]{.63\textwidth}
\vspace{0pt}\centering
    \resizebox{!}{\figureheight}{
\begin{tikzpicture}[
        block/.style={draw, fill=white, rectangle, minimum width=3cm,rounded corners=0.2cm},
        trainableblock/.style={block,fill=blue!10},
        fnblock/.style={block,fill=green!10},
        operation/.style={draw, fill=red!10, circle, minimum width=2em},
        tensor/.style={draw, fill=VioletRed!50, circle},
        input/.style={block,fill=WildStrawberry!40},
    ]

    \pgfdeclarelayer{arrowlayer}
    \pgfdeclarelayer{residuallayer}
    \pgfdeclarelayer{decorations}
    \pgfdeclarelayer{bg}
    \pgfsetlayers{bg,arrowlayer,residuallayer,main,decorations}
    \node [input] (conditioning_s) {$s$};
    \node [fnblock, below=0.5cm of conditioning_s] (s_enc) {Sinusoidal Encoding};
    \node [trainableblock, below=0.2cm of s_enc] (s_emb) {MLP};

    \node [input, right=1cm of conditioning_s] (conditioning_u) {$u$};
    \node [fnblock, below=0.5cm of conditioning_u] (u_enc) {Sinusoidal Encoding};
    \node [trainableblock, below=0.2cm of u_enc] (u_emb) {MLP};
    \node [trainableblock, below=1.5cm of s_emb] (t_emb) {MLP};
    \node [fnblock, below=0.2cm of t_emb] (t_enc) {Sinusoidal Encoding};
    \node [input, below=0.5cm of t_enc] (timestep) {Timestep};

    \node [operation, below=0.5cm of u_emb] (y_plus) {$+$};
    \node [tensor, below=0.5cm of y_plus] (y_final) {y};

    \begin{pgfonlayer}{arrowlayer}
        \draw [->] (timestep) -- (t_emb) |- (y_plus);
        \draw [->] (conditioning_s) -- (s_emb) |- (y_plus);
        \draw [->] (conditioning_u) -- (u_emb) -- (y_plus);
        \draw [->] (y_plus) -- (y_final);
    \end{pgfonlayer}

    \node [input, right=2.5cm of conditioning_u, align=center] (image) {Noised Normalized\\Latent};

    \node [fnblock, below=0.5cm of image] (patching) {Patching};
    \node [trainableblock, below=0.2cm of patching] (patch_proj) {Linear};
    \node [operation, below=0.5cm of patch_proj] (image_plus_pos) {$+$};
    \node [trainableblock, left=0.2cm of image_plus_pos,align=center] (pos_emb) {Positional\\ Embedding};

    \node [tensor, below=0.5cm of image_plus_pos] (x_in) {$x$};
    \begin{pgfonlayer}{arrowlayer}
        \draw [->] (image) -- (image_plus_pos);
        \draw [->] (pos_emb) -- (image_plus_pos);
        \draw [->] (image_plus_pos) -- (x_in);
    \end{pgfonlayer}

    \node [trainableblock, below=1.5cm of $(y_final)!0.5!(x_in)$, minimum width=5cm] (attn1) {SingleStream-DiT-Block 1};
    \node [trainableblock, below=0.3cm of attn1, minimum width=5cm] (attn2) {SingleStream-DiT-Block 2};
    \node [below=0.4cm of attn2, minimum width=5cm] {$\ldots$};
    \node [trainableblock, below=1.3cm of attn2, minimum width=5cm] (attnN) {SingleStream-DiT-Block N};

    \node [trainableblock, below=1.5cm of attnN.north] (out_mod) {Modulation};
    \node [trainableblock, below=0.2cm of out_mod] (out_mlp) {Linear};
    \node [fnblock, below=0.2cm of out_mlp] (depatch) {Unpatching};
    \node [input, below=0.5cm of depatch] (output) {Output};

    \coordinate (attn1_left) at ($(attn1.west)+(-2, 0)$);

    \begin{pgfonlayer}{arrowlayer}
        \draw [->] (x_in) -| (attn1.north) -- (output);

        \draw [->] (y_final) -| (attn1_left) -- (attn1);
        \draw [->] (attn1_left) |- (attn2);
        \draw [->] (attn1_left) |- (attnN);
        \draw [->] (attn1_left) |- (out_mod);
    \end{pgfonlayer}

    \begin{pgfonlayer}{bg}
        \draw [fill=blue!5] ($(attn1.north east)+(1, 0.5)$) -| ($(attnN.south west)+(-1, -0.5)$) -| cycle;
    \end{pgfonlayer}

\end{tikzpicture}
}

    \caption{Model architecture derived from SD3.     \label{fig:model_archi}}
\end{minipage}\hfill
\begin{minipage}[t]{.33\textwidth}
\vspace{0pt}\centering

    \resizebox{!}{\figureheight}{\begin{tikzpicture}[
        block/.style={draw, fill=white, rectangle, minimum width=3cm, rounded corners=0.2cm},
        trainableblock/.style={block,fill=blue!10},
        fnblock/.style={block,fill=green!10},
        operation/.style={draw, fill=red!10, circle, minimum width=2em},
        tensor/.style={draw, fill=VioletRed!50, circle},
    ]

    \pgfdeclarelayer{arrowlayer}
    \pgfdeclarelayer{residuallayer}
    \pgfsetlayers{arrowlayer,residuallayer,main}

    \newcommand{\inblockdistance}{0.2cm}
    \newcommand{\outblockdistance}{0.5cm}
    \newcommand{\leftmodpos}{-3.2cm}

    \node [tensor] (x) {$x$};

    \node [below=\outblockdistance of x, fnblock] (x_ln1) {Layernorm};
    \node [below=\inblockdistance of x_ln1, fnblock] (x_mod) {Mod: $\alpha \cdot \bullet + \beta$};
    \node [below=\inblockdistance of x_mod, trainableblock] (x_lin) {Linear};

    \node [below=0.7cm of x_lin, fnblock, minimum width=4.5cm, minimum height=1.5cm, align=center] (attn) {\vphantom{Attention}\\ \Large Attention};
    \node [below=0.1cm of attn.north] (k) {\Large \emph{K}};
    \node [left= 0.7cm of k.center] (q) {\Large \emph{Q}};
    \node [right=0.7cm of k.center] (v) {\Large \emph{V}};

    \coordinate (x_lin_qout) at (x_lin.south -| q.center);
    \coordinate (x_lin_kout) at (x_lin.south -| k.center);
    \coordinate (x_lin_vout) at (x_lin.south -| v.center);

    \begin{pgfonlayer}{arrowlayer}
        \draw [->] (x_lin_qout) -- (q |- attn.north);
        \draw [->] (x_lin_kout) -- (k |- attn.north);
        \draw [->] (x_lin_vout) -- (v |- attn.north);
    \end{pgfonlayer}

    \node [below=0.7cm of attn, trainableblock] (x_proj) {Linear};
    \node [below=\inblockdistance of x_proj, operation] (x_proj_gated) {$*$};
    \node [below=\outblockdistance of x_proj_gated, operation] (x_attn_out) {$+$};
    \node [below=\outblockdistance of x_attn_out, fnblock] (x_ln2) {Layernorm};
    \node [below=\inblockdistance of x_ln2, fnblock] (x_mod2) {Mod: $\delta \cdot \bullet + \epsilon$};
    \node [below=\inblockdistance of x_mod2, trainableblock] (x_mlp) {MLP};
    \node [below=\inblockdistance of x_mlp, operation] (x_mlp_gated) {$*$};
    \node [below=\outblockdistance of x_mlp_gated, operation] (x_out) {$+$};
    \path let \p1=($(x_mod.north west)!0.5!(x_mod.west)+(0, 0.25)$) in node [tensor] at (\leftmodpos, \y1) (alpha_x) {$\scriptstyle \alpha$};
    \path let \p1=($(x_mod.south west)!0.5!(x_mod.west)-(0, 0.25)$) in node [tensor] at (\leftmodpos, \y1) (beta_x) {$\scriptstyle \beta$};
    \path let \p1=(x_proj_gated.west) in node [tensor] at (\leftmodpos, \y1) (gamma_x) {$\scriptstyle \gamma$};
    \path let \p1=($(x_mod2.north west)!0.5!(x_mod2.west)+(0, 0.25)$) in node [tensor] at (\leftmodpos, \y1) (delta_x) {$\scriptstyle \delta$};
    \path let \p1=($(x_mod2.south west)!0.5!(x_mod2.west)-(0, 0.25)$) in node [tensor] at (\leftmodpos, \y1) (epsilon_x) {$\scriptstyle \epsilon$};
    \path let \p1=(x_mlp_gated.west) in node [tensor] at (\leftmodpos, \y1) (zeta_x) {$\scriptstyle \zeta$};

    \node [tensor, above left=2cm and 2.5cm of x] (y) {$y$};
    \node [fnblock, below=\outblockdistance of y] (y_silu) {$\mathrm{SiLU}$};
    \node [trainableblock, below=\inblockdistance of y_silu] (y_linear) {Linear};
    \coordinate (y_split) at ($(y_linear.south)+(-1, -0.3)$);

    \coordinate (x_attn_out_left) at ($(x_attn_out) + (-2.5, 0)$);

    \begin{pgfonlayer}{arrowlayer}
        \draw [->] (x) -- (x_lin);
        \draw [->] (attn) -- (x_proj);
        \draw [->] (x_proj) -- (x_attn_out);
        \draw [->] (x_attn_out) -- (x_out);
        \draw [<-] (y.north) --++ (0, 0.4);
        \draw (y) -- (y_linear) |- (y_split);
        \draw [->] (y_split) |- (alpha_x);
        \draw [->] (y_split) |- (beta_x);
        \draw [->] (y_split) |- (gamma_x);
        \draw [->] (y_split) |- (delta_x);
        \draw [->] (y_split) |- (epsilon_x);
        \draw [->] (y_split) |- (zeta_x);

        \draw [->] (alpha_x) -- ($(x_mod.north west)!0.5!(x_mod.west)$);
        \draw [->] (beta_x) -- ($(x_mod.south west)!0.5!(x_mod.west)$);
        \draw [->] (gamma_x) -- (x_proj_gated.west);
        \draw [->] (delta_x) -- ($(x_mod2.north west)!0.5!(x_mod2.west)$);
        \draw [->] (epsilon_x) -- ($(x_mod2.south west)!0.5!(x_mod2.west)$);
        \draw [->] (zeta_x) -- (x_mlp_gated.west);

    \end{pgfonlayer}
    \begin{pgfonlayer}{residuallayer}
        \draw [->, line width=2mm, draw=white] (x) -| (x_attn_out_left) -- (x_attn_out);
        \draw [->, line width=2mm, draw=white] (x_attn_out_left) |- (x_out);
        \draw [<-, thick] (x) --++ (0, 3.1);
        \draw [->, thick] (x) -| (x_attn_out_left) -- (x_attn_out);
        \draw [->, thick] (x_attn_out_left) |- (x_out);
        \draw [->, thick] (x_out) --++ (0, -1);
    \end{pgfonlayer}
\end{tikzpicture}}
    \caption{SingleStream-DiT block architecture.  *  denotes element-wise multiplication.    \label{fig:single_stream_block}}
\end{minipage}
\end{figure}

\subsubsection{Conditioning via MultiControlNet}
To incorporate auxiliary modalities (DSM and Pléiades RGB imagery), we adopt a multi-branch ControlNet~\cite{zhang_adding_2023} strategy. Specifically, we instantiate two distinct ControlNets: one for the DSM modality and one for the Pléiades RGB modality. At each transformer block, the modality-specific residuals are added element-wise before injection into the pruned SD3 backbone. We use this element-wise residual sum following the ControlNet formulation~\cite{zhang_adding_2023}; other fusion mechanisms were not ablated and are left for future work.

\subsection{Data}
\label{sec:data}

We assembled a multi-city dataset from LiDAR-HD, orthorectified Pléiades-HR imagery, and CARS stereo DSMs. The paired multimodal subset covers 9 French cities; raw LiDAR coverage spans 20 cities, while the LiDAR-only backbone training split uses the 19 non-held-out cities. All products were processed as $512\times512$ pixel patches at 0.5~m~resolution.

\subsubsection{Data Acquisition and Preprocessing}

We obtained LiDAR-HD point clouds from the IGN website (\url{https://cartes.gouv.fr/telechargement/IGNF\_NUAGES-DE-POINTS-LIDAR-HD}, accessed on 1 December 2025) and generated 0.5\,m elevation rasters by median aggregation with CARS-rasterize (\url{https://github.com/CNES/cars-rasterize}, accessed on 1 December 2025), a CARS~\cite{youssefi_cars_2020} plugin. The dataset-generating workflow fills connected LiDAR NoData components below a sieve threshold of 800 pixels (200\,m$^2$ at 0.5\,m resolution). It processes $2048\times2048$ pixel blocks (1024\,m~$\times$~1024\,m), with 50-pixel overlap, and iteratively fills eligible pixels from a radius-3 ($7\times7$ pixel) neighborhood using the mean of values below the local 40th percentile, for at most five iterations. The percentile and neighborhood choices were empirical choices informed by tests around buildings, not a theoretically optimized or sensitivity-tested rule. A $512\times512$ target crop containing any remaining LiDAR NaN is discarded. The dataset is primarily urban and includes LiDAR-HD data from 20 French cities: Amiens, Angers, Arcachon, Biarritz, Bordeaux, Clermont-Ferrand, Grenoble, Lyon, Marseille, Montpellier, Nancy, Nantes, Nice, Paris, Poitiers, Reims, Rennes, St-Etienne, Strasbourg, and Toulouse.

The stereo DSMs were supplied already processed by the CARS workflow~\cite{youssefi_cars_2020}. These DSMs have a spatial resolution of 0.5\,m, matching that of the input Pléiades-HR imagery. The sensor imagery was delivered with the following geometric settings:
\begin{itemize}
    \item Geometric processing: sensor;
    \item Ephemeris used: corrected;
    \item Attitudes used: accurate;
    \item Ground setting: true;
    \item Ground description: \texttt{R3D\_ORTHO};
    \item Vertical setting: false.
\end{itemize}

Orthorectification through Reference3D (part of the Airbus DS Elevation30 suite) and reprojection to Lambert93 provide the geometric starting point for comparison, but do not by themselves establish vertical co-registration. For each DSM acquisition, we estimated an acquisition-level offset from the overlap with LiDAR by forming a global histogram of valid DSM--LiDAR differences over $\pm100$\,m with 4096 equal-width bins and subtracting the modal bin center. The same histogram settings were used for all cities, including Bordeaux. This co-registration is required to evaluate local refinement: without it, the errors of both the input DSM and the improved DSM would contain an acquisition-wide vertical offset that obscures the effect of local corrections. Estimating such an absolute vertical offset from DSM and RGB alone is not the intended task and is generally impractical without an external vertical anchor; the model therefore assumes a vertically calibrated DSM input (assumption A1, Section~\ref{sec:task_assumptions}).

Our conditional task also assumes that the DSM and LiDAR represent a compatible underlying surface (assumption A2, Section~\ref{sec:task_assumptions}). This requirement may not be met when the physical surface differs between acquisitions, for instance due to construction, demolition, or seasonal vegetation state. We therefore retain DSM patches whose calibrated RMSE is below the city-specific 90th percentile. The LiDAR target remains available when DSM conditioning is omitted. The choice of the 90th percentile was made empirically during dataset construction as a practical trade-off between pair compatibility and data quantity; it was not ablated or optimized. Appendix~\ref{app:data_support} reports the curation counts, and Appendix~\ref{app:qualitative} illustrates representative excluded cases. Developing more sophisticated quality-control pipelines to identify high-quality DSM--RGB--LiDAR pairs is left for future work.

The finite high-RMSE excluded patches define a separate curation cohort of 4064~geographic patches. Eighteen of these patches have no Pléiades acquisition listed (17 in Montpellier and one in Bordeaux), so the common DSM-only/DSM~+~RGB stress evaluation in Appendix~\ref{app:evaluation} uses the remaining 4046 RGB-supported patches.

The Pléiades RGB imagery was also provided orthorectified, albeit without a common radiometric calibration. When pansharpened imagery was not supplied by Airbus, we performed pansharpening using GDAL~\cite{rouault_gdal_2026}. For each acquisition and RGB band, image intensities were clipped using the 2nd and 98th percentiles. We discarded the NIR band because the SD3 VAE accepts only three input channels. Where available, multiple temporal acquisitions over the same location were included. They remain grouped in one geographic sample, and one acquisition is drawn randomly during collation, so additional dates do not increase that location's sampling weight.

Pléiades-HR imagery and DSMs were obtained for 9 French cities: Amiens, Arcachon, Biarritz, Bordeaux, Montpellier, Nice, Paris, Strasbourg, and Toulouse. Table~\ref{tab:dataset_stats} summarizes modality availability by city, distinguishing geographic patch locations from Pléiades acquisition--patch pairs.

\begin{table}[!htbp]
\small
    \caption{Dataset constitution by city and modality in terms of $512\times512$ geographic patches. \textbf{Pléiades} counts locations with at least one acquisition, whereas \textbf{All Pléiades} counts acquisition--patch pairs across all dates.}
    \label{tab:dataset_stats}

    \begin{tabularx}{\textwidth}{lRRRR}
        \toprule
        \textbf{City} & \textbf{LiDAR} & \textbf{Pléiades} & \textbf{All Pléiades} & \textbf{Stereo DSM} \\
        \midrule
        Amiens            &  27,645 &  20,404 &  44,220 &  5077 \\
        Angers            &   5380 &      0 &      0 &     0 \\
        Arcachon          &  10,076 &   7132 &  21,515 &  1462 \\
        Biarritz          &  11,090 &   6571 &  23,299 &  4180 \\
        Bordeaux          &  26,191 &  14,376 &  65,273 &  3221 \\
        Clermont-Ferrand  &   9311 &      0 &      0 &     0 \\
        Grenoble          &   8379 &      0 &      0 &     0 \\
        Lyon              &  10,104 &      0 &      0 &     0 \\
        Marseille         &   7288 &      0 &      0 &     0 \\
        Montpellier       &  50,281 &  13,793 &  32,201 &  7632 \\
        Nancy             &   9111 &      0 &      0 &     0 \\
        Nantes            &  10,556 &      0 &      0 &     0 \\
        Nice              &  18,454 &  12,136 &  41,165 &  2200 \\
        Paris             &  45,382 &  20,764 &  66,433 &  4752 \\
        Poitiers          &   7122 &      0 &      0 &     0 \\
        Reims             &   6102 &      0 &      0 &     0 \\
        Rennes            &   7872 &      0 &      0 &     0 \\
        St-Etienne        &   6174 &      0 &      0 &     0 \\
        Strasbourg        &  53,679 &  35,084 &  71,623 &  3559 \\
        Toulouse          &  32,549 &  20,848 &  46,246 &  4442 \\
        \midrule
        Total             & 362,746 & 151,108 & 411,975 & 36,525 \\
        \bottomrule
    \end{tabularx}
\end{table}

\subsubsection{Spatial Alignment and Partitioning}
To place the rasters on a common evaluation grid, all rasters were reprojected into the Lambert93 coordinate system at a 0.5\,m resolution using GDAL and GNU Parallel~\cite{tange_gnu_2018}. This allows sliding-window cropping directly in pixel space, although residual geometric misalignment may remain.

Patch extraction precedes splitting. The builder scans full non-overlapping $512\times512$ windows with stride 512 and drops border remainders. Each $(\text{city},x,y)$ location defines one geographic patch; adjacent patches may touch but share no pixels, and no geographic buffer is applied. All Pléiades acquisitions attached to one location remain grouped in the same split.

The split depends on the training stage. The LiDAR training set is used for backbone adaptation. Locations with LiDAR+Pléiades but no DSM conditioning are divided into approximately 90\% training and 10\% validation sets, while fully paired LiDAR+DSM+Pléiades patches are divided into 70\% training, 10\% validation, and 20\% test sets within each non-held-out city. All fully paired Bordeaux patches are reserved for the held-out test city; Bordeaux locations without all three modalities are not used for training or validation. No geographic patch appears in more than one of the training, validation, in-context test, or Bordeaux test partitions. A training or validation patch may, however, be reused in the corresponding partition across successive stages when additional modalities are available.

\begin{figure}[!htbp]
\centering

\begin{tikzpicture}[
    font=\sffamily\small,
    box/.style={draw=black!70, thick, align=center, inner sep=2mm, rounded corners=2pt},
    train_L1/.style={box, fill=mdpiyellow_light},
    train_L2/.style={box, fill=mdpiyellow_med},
    train_L3/.style={box, fill=mdpidarkyellow},
    valbg/.style={box, fill=mdpiblue},
    valinner/.style={box, fill=mdpidarkblue},
    testbg/.style={box, fill=mdpired}
]
\definecolor{mdpiyellow_light}{RGB}{255, 250, 230}
\definecolor{mdpiyellow_med}{RGB}{255, 242, 204}
\definecolor{mdpidarkyellow}{RGB}{255, 229, 153}
\definecolor{mdpiblue}{RGB}{218, 232, 252}
\definecolor{mdpidarkblue}{RGB}{186, 216, 250}
\definecolor{mdpired}{RGB}{248, 206, 204}
\node[train_L1, minimum width=13cm, minimum height=4.5cm] (train1) {};
\node[anchor=west, xshift=0.3cm, align=center] at (train1.west) {LiDAR\\Train};
\node[train_L2, minimum width=9.5cm, minimum height=3.8cm, anchor=east] (train2) at ([xshift=-0.3cm]train1.east) {};
\node[anchor=west, xshift=0.3cm, align=center] at (train2.west) {LiDAR + Pléiades\\Train};
\node[train_L3, minimum width=5cm, minimum height=3cm, anchor=east] (train3) at ([xshift=-0.3cm]train2.east) {LiDAR + Stereo DSM +\\Pléiades\\Train};
\node[valbg, minimum width=9.5cm, minimum height=2.2cm, anchor=south east] (val1) at ([yshift=0.3cm]train2.east |- train1.north) {};
\node[anchor=west, xshift=0.3cm, align=center] at (val1.west) {LiDAR + Pléiades\\Val};

\node[valinner, minimum width=5cm, minimum height=1.4cm, anchor=east] (val2) at ([xshift=-0.3cm]val1.east) {LiDAR + Stereo DSM\\+ Pléiades\\Val};
\node[testbg, minimum width=5cm, minimum height=2cm, anchor=south east] (test_dsm) at ([yshift=0.3cm]val2.east |- val1.north) {LiDAR + Stereo DSM\\+ Pléiades\\Test};

\end{tikzpicture}
\caption{Diagram of the partitioning strategy across modality-specific datasets. Patches from the same geographic location remain in the corresponding partition as modalities are added; all fully paired Bordeaux patches form the held-out test set. Adjacent patches may touch, and no geographic buffer is used.}
\label{fig:spatial_partitioning}
\end{figure}
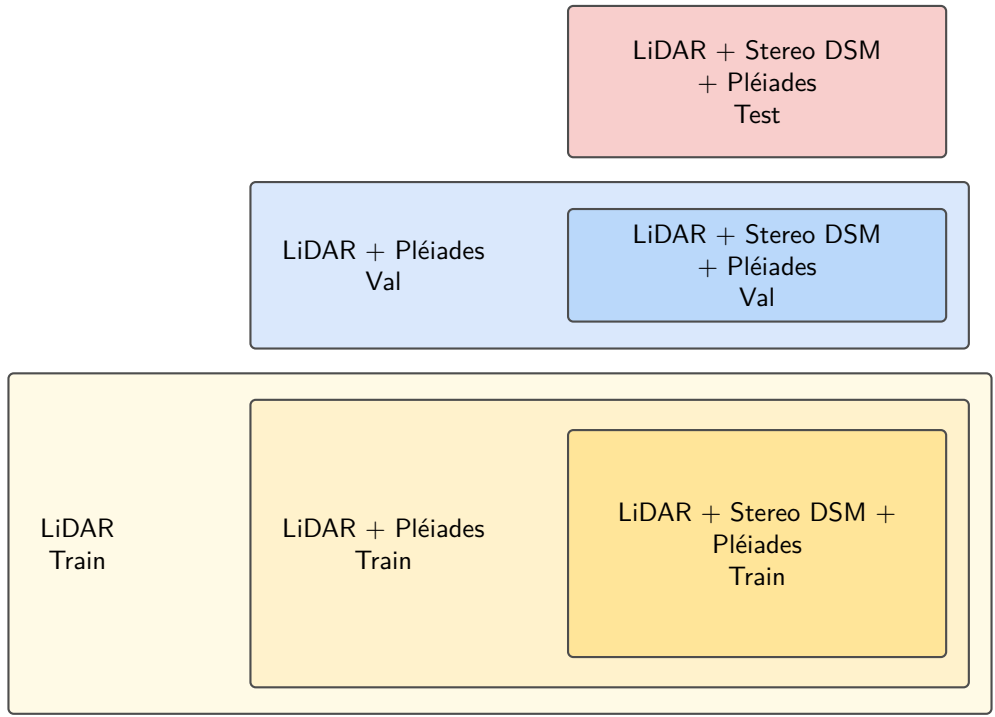

The geographic grouping is illustrated in Figure~\ref{fig:spatial_partitioning}, and acquisition dates are summarized in Figure~\ref{fig:data_timeline}. Exact split counts by city and training stage are reported in Appendix~\ref{app:data_support}, Table~\ref{tab:app_final_manifest_counts}.

\begin{figure}[!htbp]
\centering

\begin{tikzpicture}
    \node[anchor=south west] (img) at (0,0)
        {\includegraphics[width=0.82\linewidth]{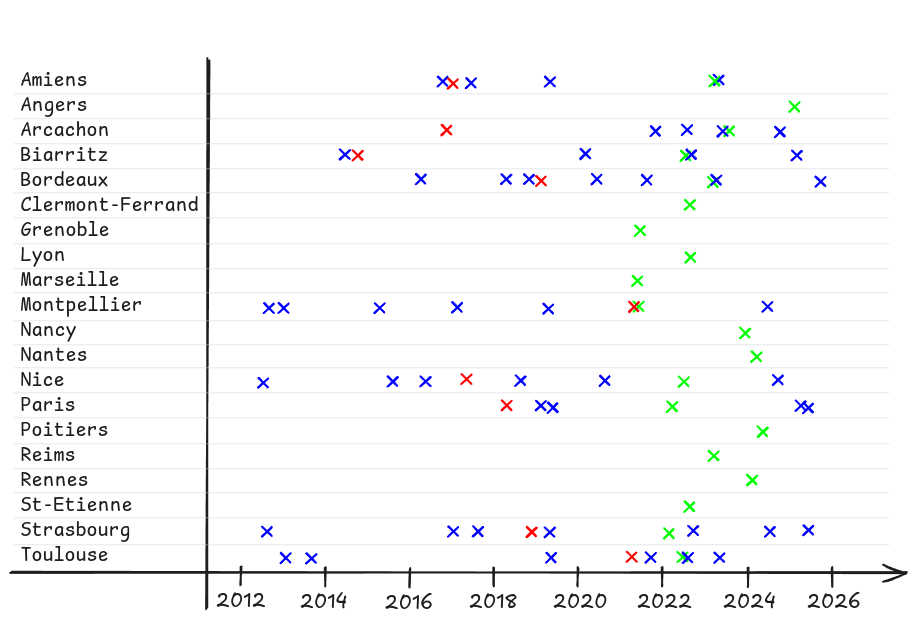}};
    \begin{scope}[x={(img.south east)}, y={(img.north west)}]

    \end{scope}
\end{tikzpicture}

\caption{Data acquisition dates by city and {modality:} \textcolor{green}{LiDAR}, \textcolor{blue}{Pléiades}, and \textcolor{red}{Stereo DSM}.}

\label{fig:data_timeline}
\end{figure}

\subsubsection{Land Cover Stratification}

To analyze model performance across heterogeneous surface types, the conditional elevation-error metrics are stratified by land-cover category. We perform a per-pixel stratification: each 0.5\,m pixel inherits its label from the Theia Land Cover 2021 map (10\,m resolution), reprojected with nearest-neighbor sampling. Consequently, one source label covers approximately $20\times20$ evaluation pixels. These labels provide a coarse stratification rather than pixel-pure object boundaries; the 10\,m interior sensitivity in Appendix~\ref{app:sensitivity} evaluates the influence of class boundaries without claiming subpixel purity. The same finite LiDAR/DSM support is used for all methods when computing the reported MAE and RMSE values.

To improve interpretability, the original Theia classes were grouped into five broader categories:

\begin{itemize}
    \item \textbf{Dense Urban:} Dense Urban;
    \item \textbf{Sparse Built-Up:} Dispersed urban, Industrial and commercial areas;
    \item \textbf{Roads:} Roads;
    \item \textbf{Croplands:} Maize, Protein crops, Rapeseed, Rice, Small-grain cereals, Soybean, Sunflower, Tubers/Roots;
    \item \textbf{Vegetation:} Deciduous forests, Coniferous forests, Orchards, Grasslands, Lawn, Heathland, Vineyards.
\end{itemize}

Table~\ref{tab:landcover_distribution} reports the number of pixels per source land-cover class and dataset split for locations containing LiDAR-HD, Pléiades-HR, and stereo DSM data.

\begin{table}[!htbp]
\small
\caption{{Millions} of pixels per Theia Land Cover 2021 class and dataset split for locations containing LiDAR-HD, Pléiades-HR, and stereo DSM data. In-context corresponds to all paired cities except Bordeaux; Bordeaux is the held-out test city.}

\label{tab:landcover_distribution}
\begin{tabularx}{\textwidth}{lRRRR}
\toprule
\textbf{Land Cover}
& \multicolumn{3}{c}{\textbf{In-Context}}
& \textbf{Bordeaux} \\
\cmidrule{2-5}

& \textbf{Train} & \textbf{Val} & \textbf{Test} & \textbf{Test} \\
\midrule
Beaches and dunes                &     27.0 &      3.0 &      5.8 &      2.4 \\
Coniferous forests               &    277.8 &     40.0 &     80.0 &     35.2 \\
Deciduous forests                &   1183.9 &    171.2 &    328.0 &    106.9 \\
Dense urban                      &    329.3 &     47.0 &     93.3 &     14.6 \\
Dispersed urban                  &   1654.4 &    241.5 &    484.6 &    441.5 \\
Glaciers and permanent snow      &      0.0 &      0.0 &      0.0 &      0.0 \\
Grasslands                       &    604.9 &     85.3 &    177.4 &    156.6 \\
Greenhouses                      &      0.0 &      0.0 &      0.0 &      0.0 \\
Heathland                        &    325.6 &     46.6 &     90.7 &      0.7 \\
Industrial and commercial areas  &     73.4 &     10.5 &     21.7 &      8.7 \\
Lawn                             &      0.4 &      0.1 &      0.2 &      0.0 \\
Maize                            &    277.1 &     35.6 &     78.7 &      8.5 \\
Mineral surfaces                 &      0.6 &      0.1 &      0.1 &      0.0 \\
Orchards                         &     61.0 &      8.7 &     17.7 &      1.6 \\
Protein crops                    &     28.5 &      4.2 &      7.5 &      0.2 \\
Rapeseed                         &     86.4 &     11.7 &     25.1 &      0.2 \\
Rice                             &      0.0 &      0.0 &      0.0 &      0.0 \\
Roads                            &    115.3 &     16.1 &     32.8 &      8.2 \\
Small-grain cereals              &    472.9 &     65.1 &    136.5 &      2.3 \\
Soybean                          &      9.5 &      1.2 &      2.5 &      1.6 \\
Sunflower                        &     36.6 &      5.4 &      9.4 &      3.4 \\
Tubers/Roots                   &    161.8 &     23.4 &     45.0 &      0.2 \\
Vineyards                        &    129.1 &     18.3 &     33.8 &     47.5 \\
Water                            &      8.5 &      1.6 &      2.9 &      0.7 \\
\bottomrule
\end{tabularx}
\end{table}

\FloatBarrier
\subsection{Training}

Training protocols differed depending on the experimental objective. We distinguish three stages: backbone adaptation, single-modality ControlNet training, and sequential multimodal ControlNet training. Common optimizer and numerical settings are reported in Appendix~\ref{app:provenance}.

\paragraph{Backbone experiments.}
For backbone comparison, the pruned SD3 architecture was trained on the LiDAR training set for 50k steps with a global batch size of 120. When initializing from pretrained SD3 weights, we used a learning rate of $1\times10^{-5}$. For training from scratch, we swept learning rates from $1\times10^{-5}$ to $4\times10^{-4}$; the best configuration used $2\times10^{-4}$.

\paragraph{DSM-only ControlNet.}
The DSM ControlNet was trained on the paired LiDAR+DSM+Pléiades training set while keeping the SD3 backbone frozen. It was first trained for 20k steps with a global batch size of 120 and a learning rate of $1\times10^{-5}$; the batch size was then increased to 360 for an additional 30k steps.

\paragraph{Multimodal conditioning.}
The two ControlNets were trained sequentially. First, the Pléiades ControlNet was trained on the LiDAR+Pléiades training set for 20k steps with a global batch size of 120, followed by 30k steps with a batch size of 360. The Pléiades ControlNet was then frozen, and the DSM ControlNet was added and trained on the paired LiDAR+DSM+Pléiades set using the same 20k + 30k-step and 120-to-360 batch schedule. Both stages used a learning rate of $1\times10^{-5}$. This is the training recipe used in our experiments; we did not conduct controlled ablations of training order, joint training, or alternative schedules.

Table~\ref{tab:training_config} summarizes the training configurations and computational cost. Training used NVIDIA A100 GPUs.

\begin{table}[!htbp]
\footnotesize
\caption{Summary of training configurations and computational cost. CN stands for ControlNet.}
\label{tab:training_config}

\begin{adjustwidth}{-\extralength}{0cm}
\begin{tabularx}{\fulllength}{lCCCCCC}
\toprule
\textbf{Experiment}
& \textbf{Trainable Modules}
& \textbf{Frozen Modules}
& \textbf{Steps}
& \textbf{Batch Size}
& \textbf{Learning Rate}
& \textbf{A100 GPU-Hours} \\
\midrule
SD3-Pruned (Finetuned)
& Backbone
& --
& 50k
& 120
& $1\times10^{-5}$
& 100 \\
SD3-Pruned (Scratch)
& Backbone
& --
& 50k
& 120
& $2\times10^{-4}$
& 100 \\
DSM-Only
& DSM CN
& Backbone
& 20k + 30k
& 120 $\rightarrow$ 360
& $1\times10^{-5}$
& 70 + 270 \\
Multimodal (Stage 1)
& Pléiades CN
& Backbone
& 20k + 30k
& 120 $\rightarrow$ 360
& $1\times10^{-5}$
& 70 + 270 \\
Multimodal (Stage 2)
& DSM CN
& Backbone, Pléiades CN
& 20k + 30k
& 120 $\rightarrow$ 360
& $1\times10^{-5}$
& 90 + 350 \\
\bottomrule
\end{tabularx}

\end{adjustwidth}
\end{table}

\FloatBarrier
\section{Results}

\subsection{Backbone}
We investigate whether elevation-map generation benefits from SD3's RGB pretraining even in this pruned architecture. To this end, we compare two configurations:
\begin{enumerate}
    \item \textbf{SD3-Pruned (Finetuned):} Our text-stream-pruned architecture keeping the pretrained image-stream weights.
    \item \textbf{SD3-Pruned (Scratch):} Our text-stream-pruned architecture trained from random initialization.
\end{enumerate}

Generation quality is evaluated through the Fréchet Distance computed over DINOv2~\cite{oquab2024dinov} embeddings from 50k images, following the evaluation approach of Stein et~al.~\cite{stein2023exposing}. Elevation maps are min--max normalized and duplicated across channels before feature extraction; generation and feature-extraction settings are reported in Appendix~\ref{app:provenance}, Table~\ref{tab:app_fd_settings}. FD$_{\mathrm{DINOv2}}$ is a relative feature-distribution distance, not an elevation error \mbox{in~meters.}

\begin{figure}[!htbp]
\centering

\resizebox{0.92\textwidth}{!}{
\begin{tikzpicture}
    \node[anchor=south west] (img) at (0,0)
        {\includegraphics[width=\linewidth]{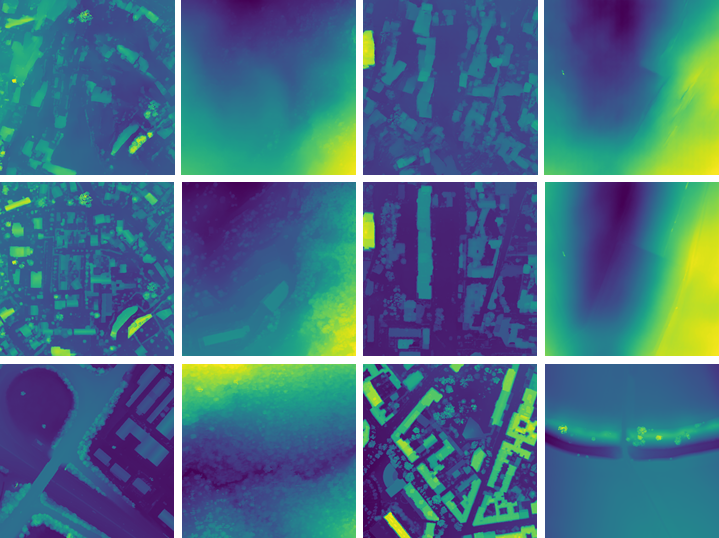}};
    \begin{scope}[x={(img.south east)}, y={(img.north west)}]
    \node[anchor=center] at (0.125,1.02) {$s$ = 3.2; $u$ = 155};
    \node[anchor=center] at (0.375,1.02) {$s$ = 28.4; $u$ = 498};
    \node[anchor=center] at (0.625,1.02) {$s$ = 10; $u$ = 43};
    \node[anchor=center] at (0.875,1.02) {$s$ = 2; $u$ = 56};
    \node[anchor=center, rotate=90, align=center] at (-0.05,0.833) {SD3-Pruned\\(Scratch)};
    \node[anchor=center, rotate=90, align=center] at (-0.05,0.5) {SD3-Pruned\\(Finetuned)};
    \node[anchor=center, rotate=90] at (-0.05,0.167) {LiDAR};
    \end{scope}
\end{tikzpicture}
}
\caption{Qualitative comparison of the two backbone configurations. From top to bottom: SD3-Pruned (Scratch), SD3-Pruned (Finetuned), and LiDAR elevation. Within each column, all generations use the same random seed. $s$ and $u$ are computed from the LiDAR patch and used to prompt the models without classifier-free guidance. Elevation maps are structural viridis min--max displays.}

\label{fig:backbone_qualitative}
\end{figure}
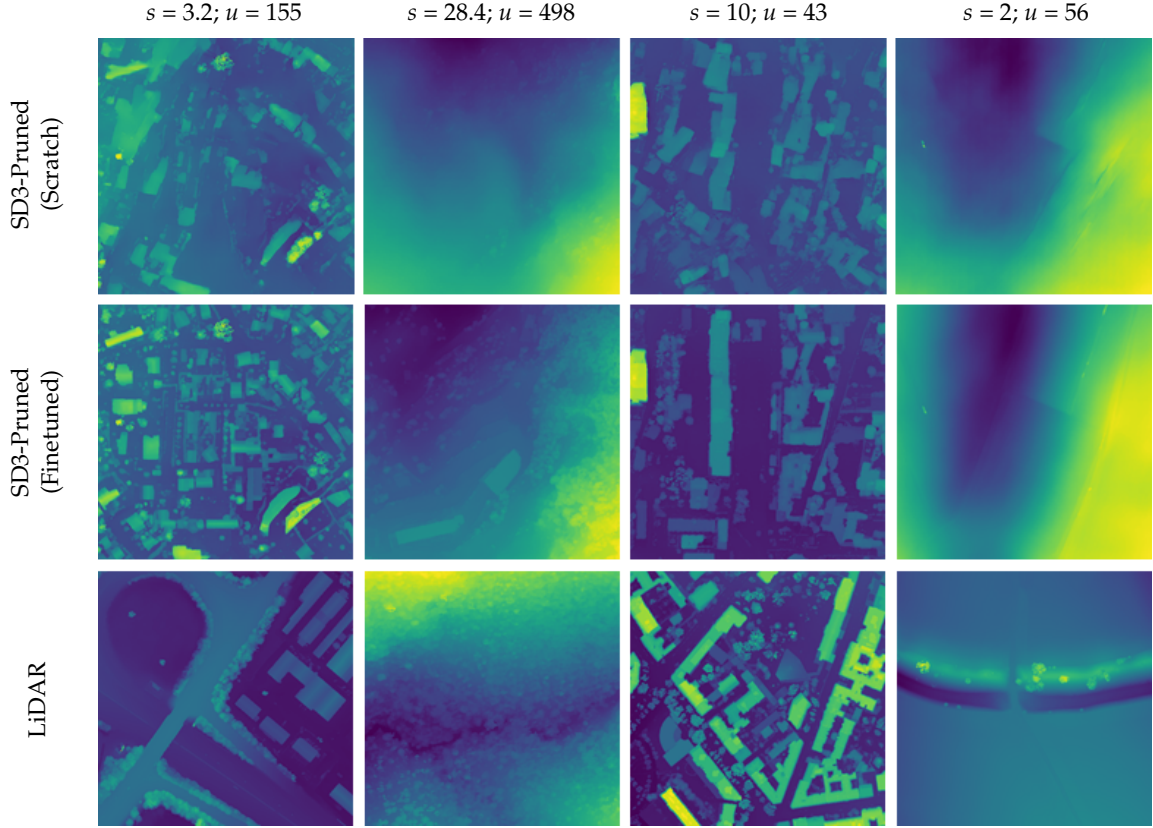

For the scratch backbone, we swept learning rates $\{1,2,5\}\times10^{-5}$ and $\{1,2,4\}\times10^{-4}$. The best scratch configuration used $2\times10^{-4}$ and obtained an FD$_{\mathrm{DINOv2}}$ of 692.8, compared with 169.4 for the pretrained configuration. Thus, pretrained initialization outperforms the best scratch configuration in the tested sweep. We did not perform a training-seed study; the numerical comparison is reported in Table~\ref{tab:ablation_backbone}. The qualitative comparison is shown in Figure~\ref{fig:backbone_qualitative}.

\begin{table}[!htbp]
\small
    \caption{Comparison of backbone initialization strategies using FD$_{\mathrm{DINOv2}}$ (lower is better). Bold indicates the best result.}

    \label{tab:ablation_backbone}
    \begin{tabularx}{\textwidth}{LC}
        \toprule
        \textbf{Model Configuration} & \textbf{FD\boldmath{$_{\mathrm{DINOv2}}$}} \boldmath{$\downarrow$} \\
        \midrule
        SD3-Pruned (pretrained) & \textbf{169.4} \\%MDPI: Please add an explanation for the use of bold  in the table footer. If the bold is unnecessary, please remove it. The following highlights are the same.
        SD3-Pruned (best scratch LR) & 692.8 \\
        \bottomrule
    \end{tabularx}
\end{table}

\subsection{Pretrained-VAE Round-Trip Diagnostic}
The 9594 LiDAR patches from the in-context and Bordeaux test sets were normalized using their own LiDAR statistics, duplicated across three channels, encoded and decoded with the pretrained SD3 VAE, and returned to meters using the same patch-specific $s,u$. This diagnostic isolates the representational distortion introduced by the VAE from errors due to the learned flow. The round trip gives meter-space bias $-0.004$ m, MAE $0.272$ m, and RMSE $0.573$ m; patch-averaged edge MAE is $0.709$ m and gradient MAE is $0.288$ m/pixel, with edge errors defined from the top 10\% of target gradient magnitudes within each patch. These quantities characterize the VAE representation rather than the learned flow model. We did not fine-tune the VAE.

\subsection{Impact of Multimodal Conditioning Across Land Covers}
\label{sec:conditioning_results}
We evaluated the effectiveness of our conditioning strategy by incrementally adding control modalities. We compare the generated elevation maps against the reference LiDAR data and the calibrated stereo DSM input. The three configurations under comparison are as follows:
\begin{itemize}
    \item \textbf{Baseline (calibrated stereo DSM input):} The unrefined, acquisition-level calibrated photogrammetric DSM, used here as a reference point.
    \item \textbf{DSM only:} The SD3-pruned backbone guided solely by the stereo DSM features.
    \item \textbf{DSM + Pléiades ControlNet:} The full multimodal setup, jointly leveraging stereo DSM and Pléiades RGB imagery.
\end{itemize}

\paragraph{Quantitative results.}

Table~\ref{tab:evaluation} reports MAE and RMSE on common finite LiDAR/DSM support, averaged over 20 inference seeds for each learned configuration. The in-context test contains 6385 geographic patches from eight cities; the held-out Bordeaux test contains 3209 patches (Appendix~\ref{app:data_support}, Table~\ref{tab:app_final_manifest_counts}). In the in-context cities, DSM-only refinement lowers RMSE in every land-cover group, although Sparse Built-Up MAE increases slightly (1.48 to 1.50\,m). Adding RGB lowers RMSE in every group relative to DSM-only; MAE also decreases except for Croplands, where it is unchanged at the displayed precision. The largest urban gain is for Dense Urban, where RMSE falls from 6.00\,m for the calibrated input to 3.85\,m with DSM-only conditioning and 3.45\,m with DSM~+~RGB.

\begin{table}[!htbp]
\scriptsize
\setlength{\tabcolsep}{3pt}
    \caption{Conditioning results on the in-context test set and held-out Bordeaux test set. Metrics are computed where both LiDAR and the input DSM are valid, then aggregated by land-cover group. Metrics are averaged over 20 inference seeds; the calibrated DSM input is deterministic. The largest inference-seed standard deviation is 0.035\,m (Appendix~\ref{app:evaluation}, Table~\ref{tab:seed_sd}); these standard deviations describe generation variability for one fitted model, including the configured RGB-acquisition selection, not spatial/test uncertainty or training-run variability.  All values are in meters (m); lower is better. Bold values indicate the lowest metric among the compared methods.}
    \label{tab:evaluation}

\begin{adjustwidth}{-\extralength}{0cm}
    \begin{tabularx}{\fulllength}{lCCCCCCCCCC}
\toprule
& \multicolumn{10}{c}{\textbf{Land Cover}} \\
\cmidrule{2-11}

\textbf{Method}

& \multicolumn{2}{c}{\textbf{Dense Urban}}
& \multicolumn{2}{c}{\textbf{Croplands}}
& \multicolumn{2}{c}{\textbf{Sparse Built-Up}}
& \multicolumn{2}{c}{\textbf{Roads}}
& \multicolumn{2}{c}{\textbf{Vegetation}} \\

\cmidrule{2-11}

& \textbf{MAE} & \textbf{RMSE}
& \textbf{MAE} & \textbf{RMSE}
& \textbf{MAE} & \textbf{RMSE}
& \textbf{MAE} & \textbf{RMSE}
& \textbf{MAE} & \textbf{RMSE} \\

\specialrule{0.08em}{0.4em}{0.3em}

\multicolumn{11}{l}{\textbf{In-context}} \\
\addlinespace[0.3em]

Calibrated stereo DSM (CARS) & 2.95 & 6.00 & 1.63 & 6.64 & 1.48 & 2.70 & 1.35 & 2.96 & 2.04 & 4.19 \\
\midrule
Ours---DSM only & 2.14 & 3.85 & \textbf{0.47} & 1.04 & 1.50 & 2.68 & 1.23 & 2.51 & 1.93 & 3.63 \\
Ours---DSM + RGB & \textbf{1.91} & \textbf{3.45} & \textbf{0.47} & \textbf{0.98} & \textbf{1.32} & \textbf{2.37} & \textbf{1.06} & \textbf{2.14} & \textbf{1.75} & \textbf{3.32} \\

\specialrule{0.08em}{0.4em}{0.3em}

\multicolumn{11}{l}{\textbf{Held-out Bordeaux}} \\
\addlinespace[0.3em]

Calibrated stereo DSM (CARS) & 2.28          & 4.16          & 0.92          & 1.92          & 1.38          & 2.39          & 1.34          & 3.12          & \textbf{1.69} & 3.12 \\
\midrule
Ours---DSM only   & 1.83 & 3.10 & 0.83 & 1.69 & 1.53 & 2.56 & 1.12 & 2.26 & 1.89 & 3.33 \\
Ours---DSM + RGB  & \textbf{1.60} & \textbf{2.77} & \textbf{0.82} & \textbf{1.61} & \textbf{1.36} & \textbf{2.34} & \textbf{1.02} & \textbf{2.07} & 1.75 & \textbf{3.10} \\

\bottomrule
\end{tabularx}

\end{adjustwidth}
\end{table}

Bordeaux shows the same clear gains for Dense Urban and Roads, but also shows that conditioning can worsen results. DSM-only conditioning increases errors for Sparse Built-Up and Vegetation relative to the calibrated input. Adding RGB recovers the Sparse Built-Up loss, whereas Vegetation MAE remains higher than the input (1.75 versus 1.69\,m) and RMSE is nearly unchanged (3.10 versus 3.12\,m). Thus, RGB generally helps the conditioned model, but does not guarantee an improvement over the input in every land-cover group.

Croplands differ strongly between the two tests. In-context input RMSE is 6.64\,m and falls to 0.98\,m with DSM~+~RGB, suggesting that a small number of large stereo errors dominate the input RMSE. In Bordeaux, input RMSE is already lower at 1.92\,m and falls more modestly to 1.61\,m. RGB adds little to Croplands MAE in either test. Homogeneous crop appearance and differences between acquisition dates are possible explanations, but were not measured separately.

\begin{figure}[!htbp]
\centering

\includegraphics[width=\textwidth]{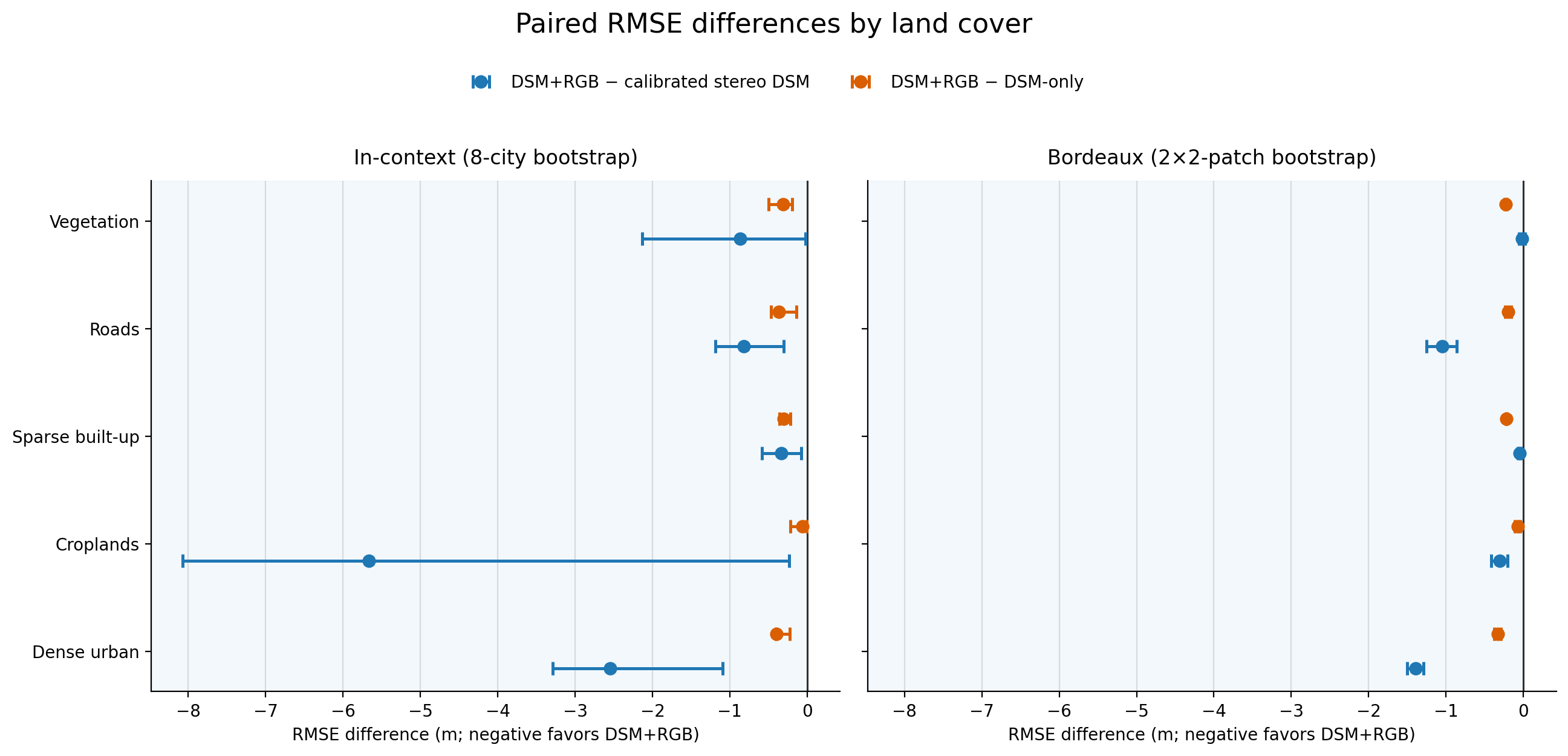}
\caption{Difference in land-cover RMSE between DSM~+~RGB and each comparator, with paired 95\% bootstrap intervals; negative values favor DSM~+~RGB. In-context intervals resample the eight paired cities, and Bordeaux intervals resample paired $2\times2$-patch spatial blocks. In every resample, RMSE is calculated for each seed 0--19 and each learned method, then averaged across seeds before the paired difference is taken. The calibrated DSM is deterministic. Thus, both comparisons use the 20-seed learned-method estimand in Table~\ref{tab:evaluation}; intervals describe spatial/test uncertainty, not inference-seed or training-run variability.}
\label{fig:landcover_rmse_ci}
\end{figure}

\paragraph{Spatial/test-sample uncertainty.}
Figure~\ref{fig:landcover_rmse_ci} shows paired RMSE differences with 95\% bootstrap intervals. Negative values favor DSM~+~RGB. Cities are resampled for the in-context test; $2\times2$-patch spatial blocks are resampled within Bordeaux. For every resample, RMSE is calculated separately for every inference seed (0--19) and learned method; the 20 RMSEs are averaged within method, and paired differences are then formed against the deterministic calibrated DSM or the 20-seed DSM-only mean. The intervals therefore quantify spatial/test-sample uncertainty for the same 20-seed estimand as Table~\ref{tab:evaluation}, with the seed set held fixed; they do not quantify training-run variability. Appendix~\ref{app:evaluation} gives the 10,000-resample protocol and separates these intervals from the inference-seed standard deviations in Table~\ref{tab:seed_sd}.

The two comparisons reveal different patterns. In context, the RMSE reduction relative to the calibrated DSM is largest for Croplands, but its city-bootstrap interval is also by far the widest. Dense Urban shows a substantial reduction, whereas the Vegetation interval approaches zero. The incremental gains over DSM-only are smaller, with Croplands closest to zero. Bordeaux has much narrower spatial-block intervals: Dense Urban and Roads show the clearest gains over the calibrated input, while Sparse Built-Up and Vegetation have differences close to zero; the Vegetation interval includes zero. Adding RGB lowers RMSE relative to DSM-only in all five Bordeaux groups, with Croplands showing the smallest benefit. The interval widths should not be compared as equivalent measures of geographic variability: the in-context bootstrap resamples cities, whereas the Bordeaux bootstrap resamples blocks within one city.

\paragraph{Complementary diagnostics.}
The headline comparison concerns valid DSM pixels. Appendix~\ref{app:void_recovery}, Table~\ref{tab:void_landcover}, evaluates original DSM voids separately, excluding pixels outside the raw DSM footprint. DSM~+~RGB has lower mean void RMSE than DSM-only in all five groups in both tests; the calibrated DSM input is undefined there. Appendix~\ref{app:evaluation}, Table~\ref{tab:stress_landcover}, additionally reports results on the P90-excluded patches. Both learned configurations reduce RMSE relative to the calibrated input in every reported group, but RGB increases Croplands RMSE relative to DSM-only (2.240 versus 2.064\,m). These distinct supports are not pooled with the headline test results. Figure~\ref{fig:app_robust_errors} complements MAE and RMSE with signed bias, NMAD, and P95 absolute error.

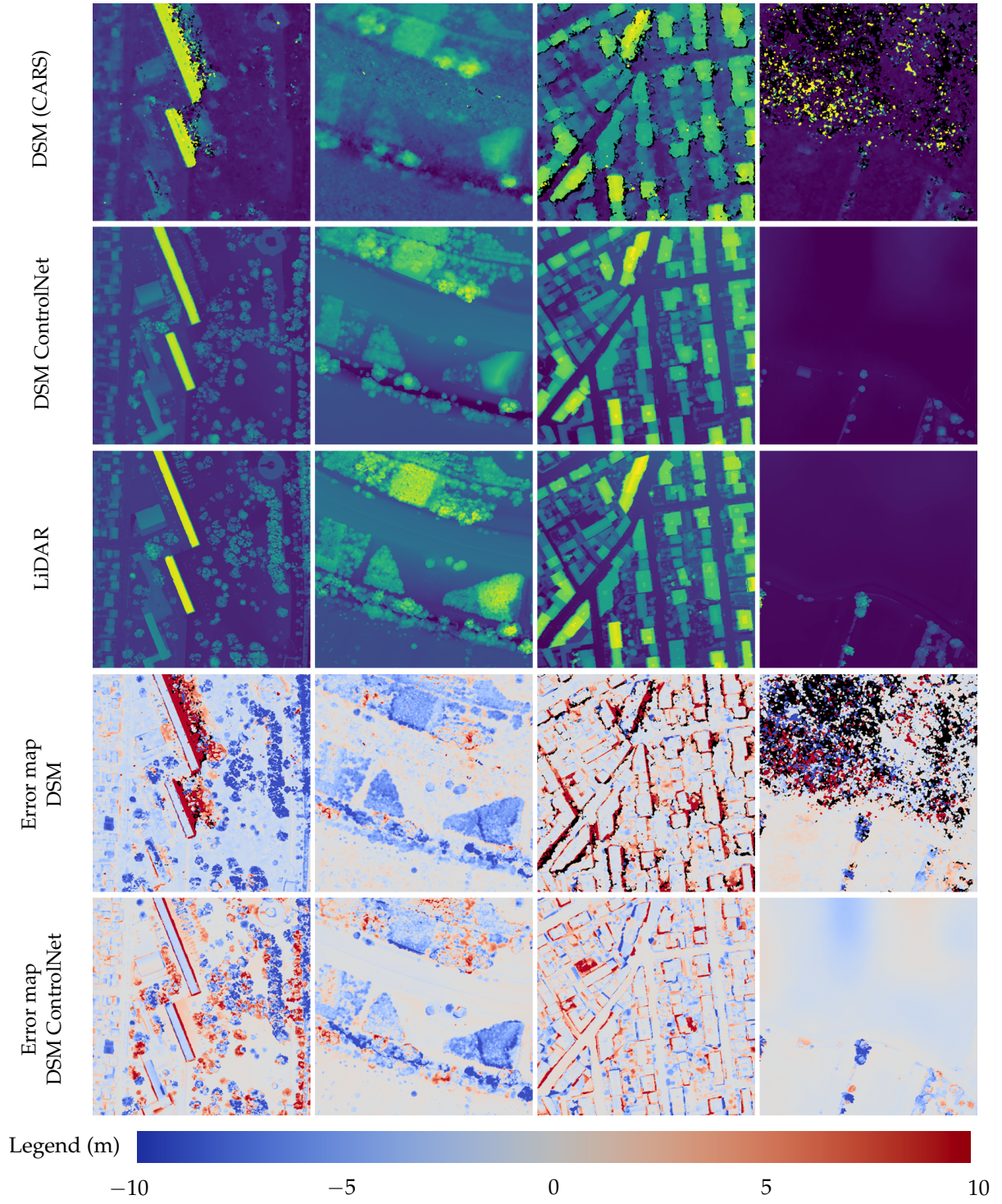
\begin{figure}[p]
\centering
\resizebox{0.94\textwidth}{!}{
\begin{tikzpicture}
    \node[anchor=south west] (img) at (0,0)
        {\includegraphics[width=\linewidth]{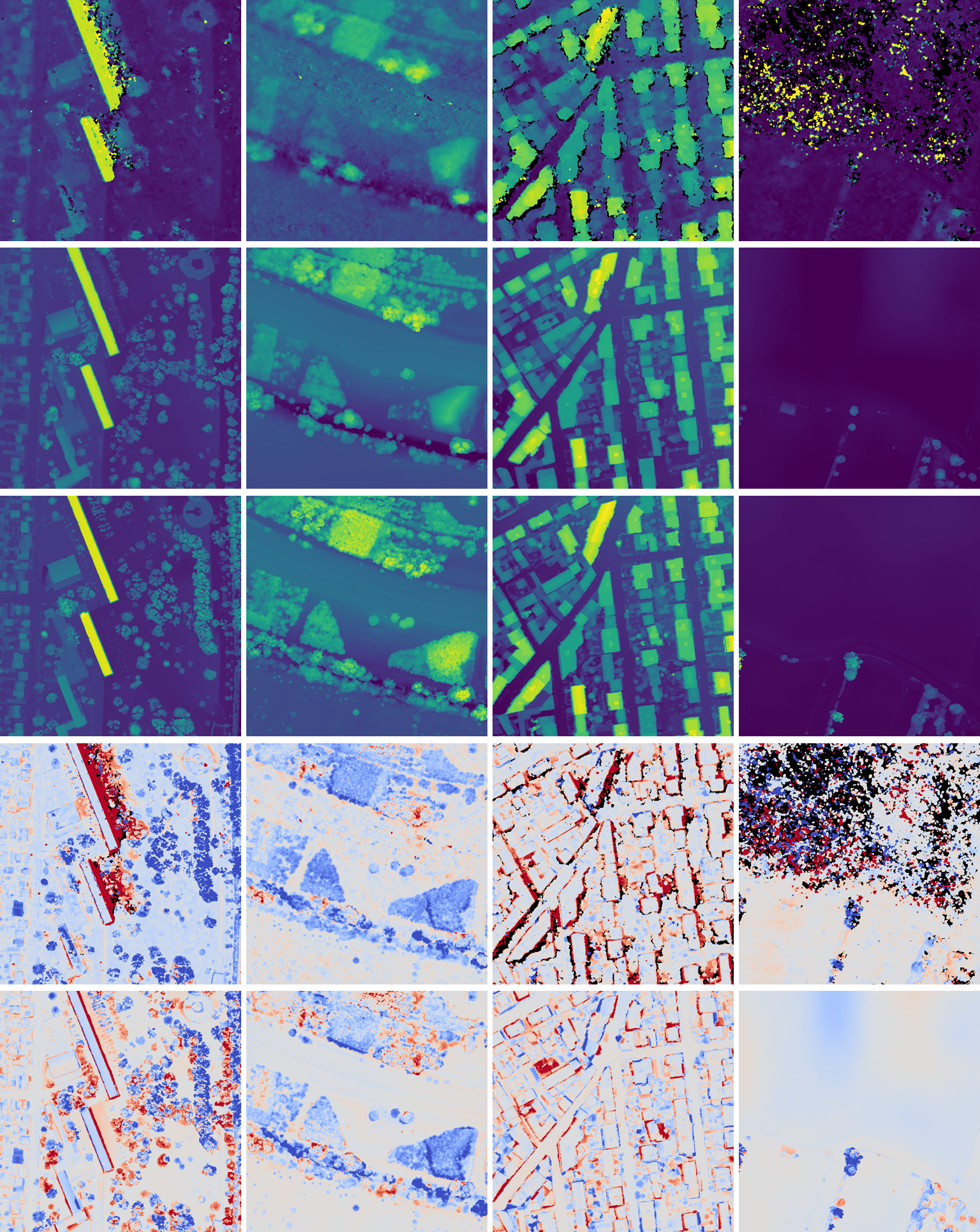}};
    \begin{scope}[x={(img.south east)}, y={(img.north west)}]
        \node[anchor=center, rotate=90] at (-0.05,0.9) {DSM (CARS)};
        \node[anchor=center, rotate=90] at (-0.05,0.7) {DSM ControlNet};
        \node[anchor=center, rotate=90] at (-0.05,0.5) {LiDAR};
        \node[anchor=center, rotate=90, align=center] at (-0.05,0.3) {Error map\\DSM};
        \node[anchor=center, rotate=90, align=center] at (-0.05,0.1) {Error map\\DSM ControlNet};
    \end{scope}
\end{tikzpicture}
}
\vspace{2pt}
\begin{tikzpicture}
    \node[anchor=east] at (0, 0) {\small Legend (m)};
    \node[anchor=west] (cbar) at (0, 0)
        {\includegraphics[width=0.8\linewidth]{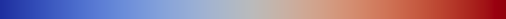}};
    \foreach \val/\pos in {-10/0, -5/0.25, 0/0.5, 5/0.75, 10/1.0}{
        \node[anchor=north] at ($(cbar.south west)!\pos!(cbar.south east)$)
            {\small $\val$};
    }
\end{tikzpicture}
\caption{DSM-only refinement. Rows show the calibrated stereo DSM, DSM-only prediction, LiDAR reference, and the corresponding errors. Within each column, elevation maps share the LiDAR range; error maps use a fixed range of $-10$ to $+10$ m.}
\label{fig:dsm_controlnet}
\end{figure}

\begin{figure}[p]
\centering
\resizebox{0.90\textwidth}{!}{
\begin{tikzpicture}
    \node[anchor=south west] (img) at (0,0)
        {\includegraphics[width=\linewidth]{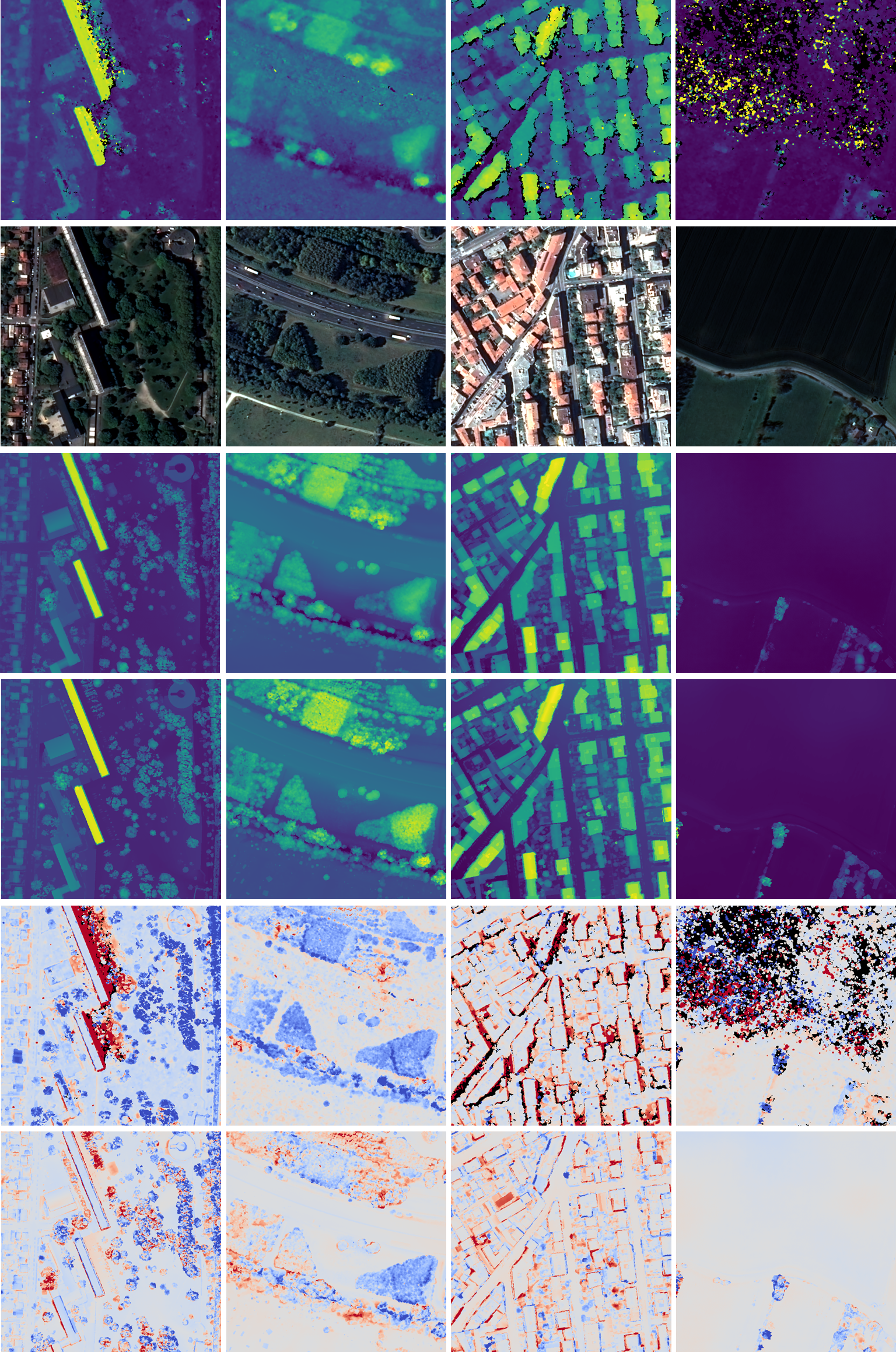}};
    \begin{scope}[x={(img.south east)}, y={(img.north west)}]
        \node[anchor=center, rotate=90] at (-0.05,0.92) {DSM (CARS)};
        \node[anchor=center, rotate=90] at (-0.05,0.75) {Pléiades};
        \node[anchor=center, rotate=90, align=center] at (-0.05,0.58) {DSM + Pléiades\\ControlNet};
        \node[anchor=center, rotate=90] at (-0.05,0.41) {LiDAR};
        \node[anchor=center, rotate=90, align=center] at (-0.05,0.25) {Error map\\DSM};
        \node[anchor=center, rotate=90, align=center] at (-0.05,0.08) {Error map\\DSM + Pléiades\\ControlNet};
    \end{scope}
\end{tikzpicture}
}
\vspace{2pt}
\begin{tikzpicture}
    \node[anchor=east] at (0, 0) {\small Legend (m)};
    \node[anchor=west] (cbar) at (0, 0)
        {\includegraphics[width=0.8\linewidth]{figs/coolwarm.png}};
    \foreach \val/\pos in {-10/0, -5/0.25, 0/0.5, 5/0.75, 10/1.0}{
        \node[anchor=north] at ($(cbar.south west)!\pos!(cbar.south east)$)
            {\small $\val$};
    }
\end{tikzpicture}
\caption{DSM~+~RGB refinement. Rows show the calibrated stereo DSM, Pléiades RGB image, multimodal prediction, LiDAR reference, and the corresponding errors. Within each column, elevation maps share the LiDAR range; error maps use a fixed range of $-10$ to $+10$ m. Pléiades ©CNES 2015/2023/2025, Distribution CNES PWH.}
\label{fig:dsm_pleiades_controlnet}
\end{figure}

\begin{figure}[p]
\centering
\resizebox{0.94\textwidth}{!}{
\begin{tikzpicture}
    \node[anchor=south west] (img) at (0,0)
        {\includegraphics[width=\linewidth]{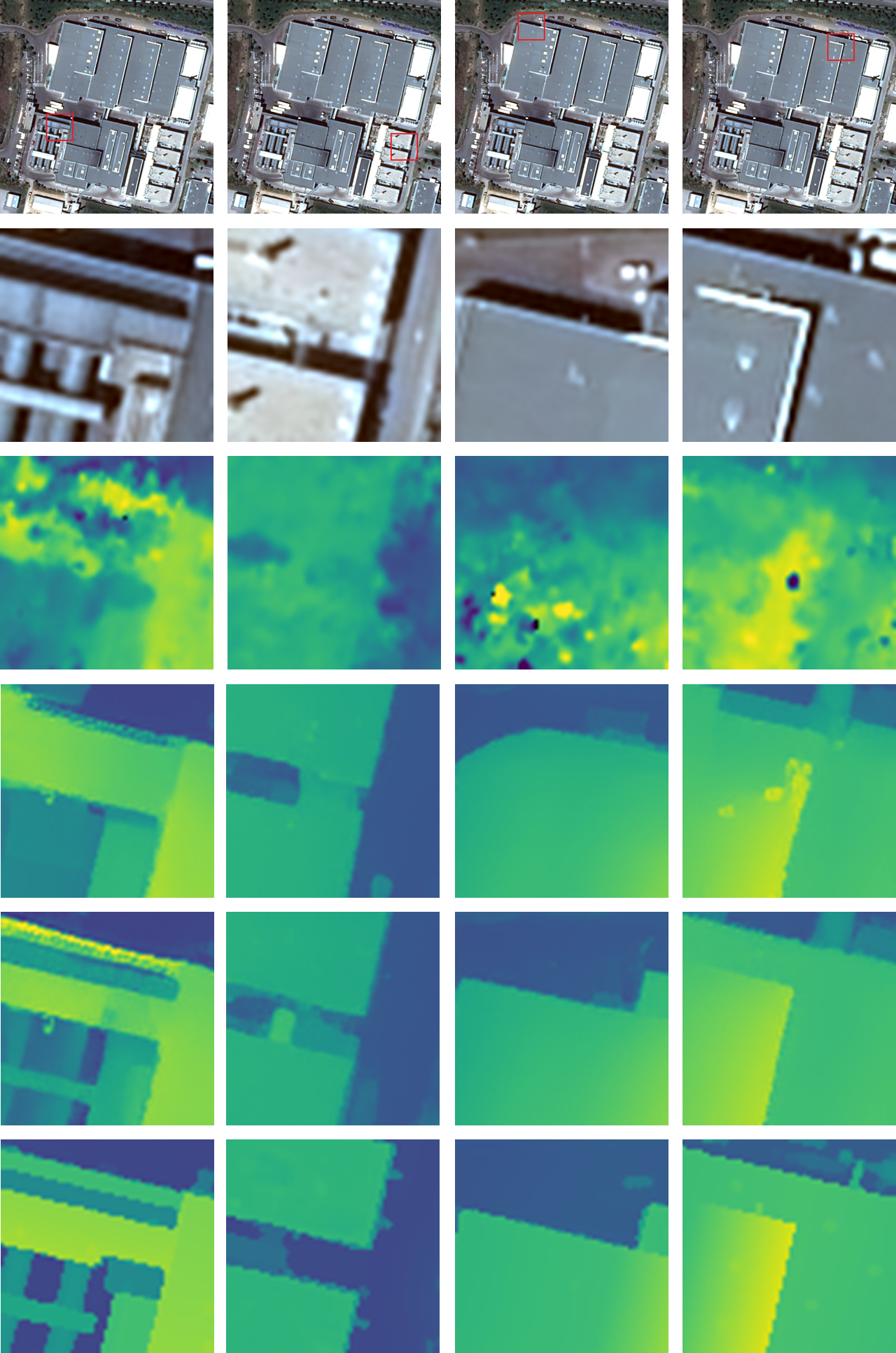}};
    \begin{scope}[x={(img.south east)}, y={(img.north west)}]
        \node[anchor=center, rotate=90] at (-0.05,0.92) {Pléiades};
        \node[anchor=center, rotate=90] at (-0.05,0.75) {Zoomed Pléiades};
        \node[anchor=center, rotate=90, align=center] at (-0.05,0.58) {Zoomed DSM\\(CARS)};
        \node[anchor=center, rotate=90, align=center] at (-0.05,0.41) {Zoomed\\DSM CN};
        \node[anchor=center, rotate=90, align=center] at (-0.05,0.25) {Zoomed DSM\\+ Pléiades CN};
        \node[anchor=center, rotate=90, align=center] at (-0.05,0.08) {Zoomed LiDAR};
    \end{scope}
\end{tikzpicture}
}
\caption{Detailed comparison of DSM-only and DSM~+~RGB refinement. Rows show the Pléiades image, enlarged image crop, calibrated stereo DSM, DSM-only prediction, DSM~+~RGB prediction, and LiDAR reference. Within each column, elevation maps share the LiDAR range. Pléiades ©CNES 2025, Distribution CNES PWH.}
\label{fig:zoom_controlnet}
\end{figure}
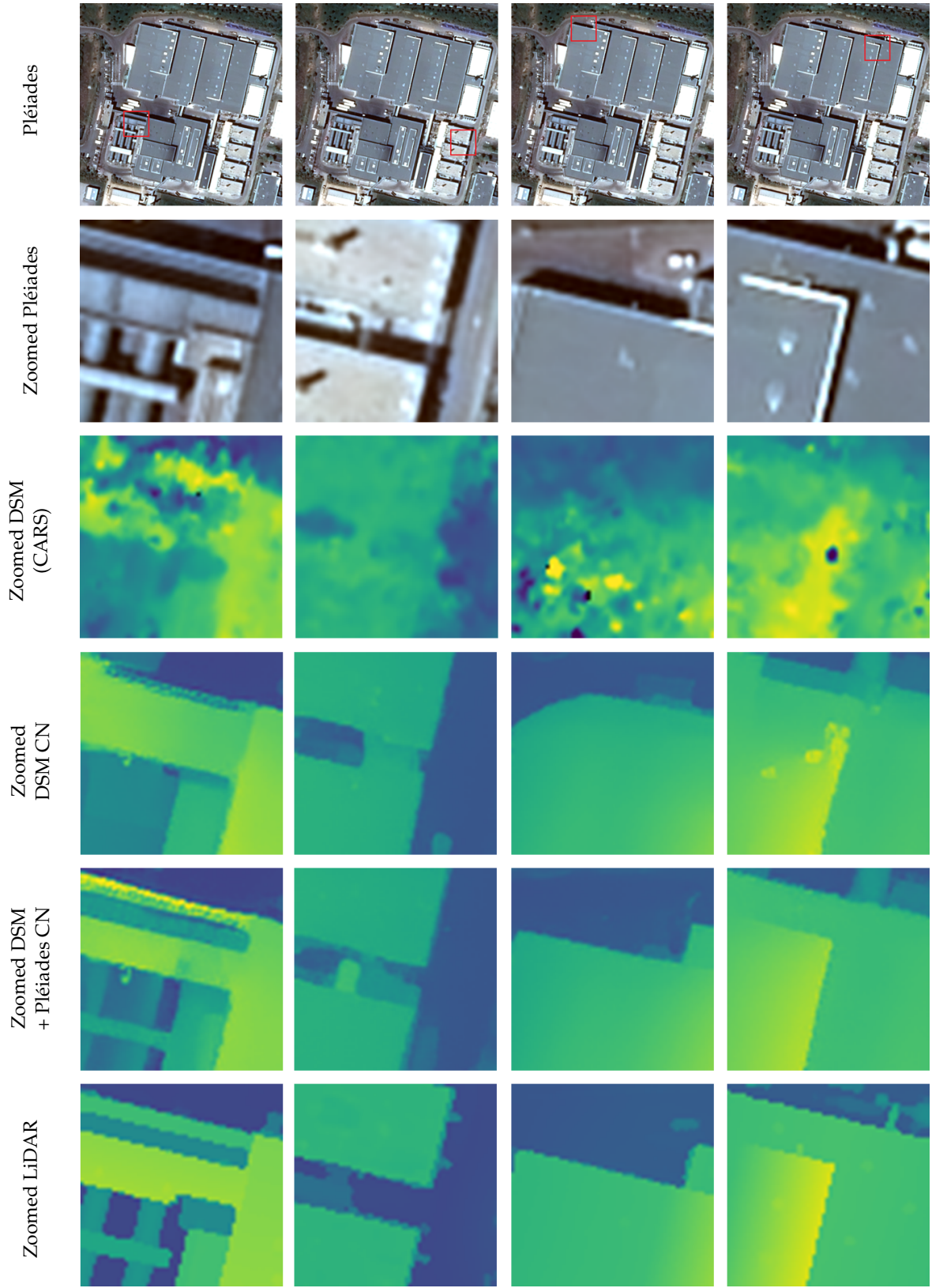

\paragraph{Qualitative analysis.}
Figures~\ref{fig:dsm_controlnet}--\ref{fig:zoom_controlnet} compare the calibrated DSM, conditioned predictions, and LiDAR reference, showing noise reduction, surface boundaries, and the placement of structures. Elevation maps share the LiDAR range within each column, and error maps use a fixed meter scale. Appendix~\ref{app:qualitative} extends this comparison with 24 metadata-random retained test patches (Figure~\ref{fig:app_random_retained}) and Bordeaux negative-transfer cases (Figure~\ref{fig:app_negative_transfer}), documenting both refinement and remaining errors.

\FloatBarrier
\section{Discussion}
The main finding is that optical imagery can complement stereo geometry when refining a photogrammetric DSM. The DSM anchors the prediction in elevation, while RGB provides image boundaries and surface cues that can help distinguish structures blurred by stereo matching. The lower errors with DSM~+~RGB, together with the sharper boundaries visible in the qualitative comparisons, support this interpretation. Dense Urban RMSE decreases by 43\% in-context and 33\% in Bordeaux relative to the calibrated input. The benefit therefore extends beyond locations in the training cities, although Bordeaux remains a single held-out city within the same acquisition and processing setting.

This complementarity is not equally useful for every surface. In Croplands, adding RGB changes MAE little on the retained tests and increases RMSE relative to DSM-only on the excluded high-error patches (Table~\ref{tab:stress_landcover}). Homogeneous appearance may provide fewer useful geometric cues, while seasonal or acquisition differences can weaken correspondence between the modalities. The large in-context Croplands gain over the calibrated input also has a wide city-bootstrap interval, indicating substantial variation between cities. Bordeaux Vegetation provides a different caution: the full model slightly worsens MAE, and its RMSE difference from the input is close to zero. Together, these results suggest that the value of RGB depends on both the input errors and the correspondence between image content and elevation, rather than on land-cover class alone.

The same combination of geometry and image cues also helps where the input DSM has no elevation. DSM~+~RGB lowers mean void RMSE relative to DSM-only in every evaluated group (Table~\ref{tab:void_landcover}). This supports void recovery as a secondary benefit of the refinement model, while the main evidence remains the correction of existing DSM elevations. The distinction matters because void pixels have no observed input value against which an improvement can be measured.

The paired bootstrap highlights the spatial heterogeneity of the gains. In particular, the large in-context Croplands reduction relative to the calibrated DSM is accompanied by a wide between-city interval. Bordeaux's narrower intervals describe within-city block variation, not transfer across cities. Table~\ref{tab:seed_sd} separately reports inference-seed standard deviations for the aggregate metrics.

The backbone results provide a broader motivation for transferring a natural-image model to elevation mapping. Pretrained initialization produces a closer LiDAR feature distribution than the best tested scratch run, and the VAE round trip shows that the image representation can retain elevation structure, albeit with measurable 0.573\,m RMSE distortion. These findings support the feasibility of the adaptation without isolating how much each component contributes to the final conditional accuracy. In particular, the feature-distance comparison is not a downstream elevation-error ablation, and the VAE reconstruction error cannot simply be subtracted from the prediction error. The DSM-only and DSM~+~RGB results should likewise be understood as a comparison of the complete conditioning configurations under the chosen training procedure.

\subsection*{Failure Modes and Limitations}
\label{sec:limitations}
The interpretation of these results depends first on the input conditions. The method assumes vertical co-registration and surface compatibility with the LiDAR reference (A1--A2, Section~\ref{sec:task_assumptions}); it does not estimate an unknown vertical datum. The RMSE filter is only an indirect compatibility heuristic, not a verified surface-change mask; it changes the composition of the evaluated data. Dense Urban pixels, for example, are more common by 1.81$\times$ among the exclusions (Table~\ref{tab:p90_composition}). Evaluation on excluded high-error patches broadens the evidence, but does not replace training and testing under alternative filtering rules. The effect of interpolating small voids in the LiDAR reference also remains unquantified around roofs and canopies.

Differences between acquisitions can reduce agreement between the conditioning inputs and the LiDAR reference. Figure~\ref{fig:app_original_stress} illustrates changes in buildings, vegetation, and shoreline appearance across RGB dates. Such changes, together with residual misalignment, complicate the interpretation of DSM--LiDAR differences. The Bordeaux examples in Figure~\ref{fig:app_negative_transfer} also document local degradation in vegetation and built areas, despite the aggregate improvements. Performance on high-rise buildings, bridges, and steep terrain remains to be evaluated separately.

Bordeaux provides evidence for one held-out French city within the same Pl\'eiades-HR/CARS setting, not other countries, sensors, terrain types, or photogrammetric pipelines. In-context patches have no pixel overlap, and RGB dates are grouped by geographic row, but adjacent training and test patches can touch without a spatial buffer (Appendix~\ref{app:data_support}). City-level bootstrap intervals do not remove this spatial dependence between splits. The 10\,m Theia labels transferred to the 0.5\,m evaluation grid also provide coarse strata rather than object-level labels. The interior-mask diagnostic in Appendix~\ref{app:sensitivity}, Figure~\ref{fig:app_boundary_sensitivity}, assesses sensitivity to excluding boundaries, while the robust-error summaries in Figure~\ref{fig:app_robust_errors} characterize bias and error tails. Eroding the masks reduces boundary mixing but cannot establish subpixel class purity.

Finally, the experimental comparisons establish gains over the calibrated DSM and the DSM-only configuration, not superiority over other enhancement methods. Task-matched conventional and supervised baselines, alternative fusion and training strategies, and independently repeated training runs remain necessary to strengthen attribution. Wider deployment would also require assessing continuity across tile boundaries. These are practical extensions of the present patch-based study rather than capabilities established by the reported tests.

\section{Conclusions}
We adapted a pretrained diffusion model to refine CARS DSMs satisfying two assumptions: vertical co-registration and surface compatibility with the LiDAR reference (A1--A2, Section~\ref{sec:task_assumptions}). In the evaluated French cities, DSM~+~RGB reduces Dense Urban RMSE from 6.00 to 3.45\,m in-context and from 4.16 to 2.77\,m in Bordeaux. Dense urban areas and roads show clear gains, while Bordeaux Vegetation shows that refinement can be neutral or slightly harmful.

These results support multimodal DSM refinement within the tested data conditions. Future work will investigate larger geographic scales, other cities and sensors, and unfiltered, independently calibrated inputs. Timestep distillation could reduce the number of sampling steps and make large-area processing more practical.

A further direction is to assess whether refining stereo DSMs acquired before and after a natural disaster, such as a landslide or earthquake, can improve elevation-change maps. This application would require checking that refinement preserves genuine changes rather than suppressing or introducing them. The adapted diffusion backbone could also be investigated as a pretrained representation for downstream geospatial tasks such as segmentation and object detection.
\vspace{6pt}

\backmatteritem{Author Contributions}{Writing---original draft, data preparation, model training, and evaluation, A.L.; supervision, writing---review and editing and data provision, S.M.; supervision and writing---review and editing, V.B. and D.D.; data preparation, B.N. All authors have read and agreed to the published version of the manuscript.}

\backmatteritem{Funding}{This research received no external funding.}

\backmatteritem{Data Availability Statement}{The LiDAR-HD, Pléiades-HR, and derived stereo-DSM imagery used in this study cannot be redistributed by the authors because of IGN/CNES and commercial-imagery licensing restrictions. With author approval, the paper includes rendered derived qualitative panels with attribution; source TIFFs, raw patches, and model weights are not redistributed. The implementation is not publicly released. Configuration details and non-restricted summaries supporting the findings are available from the corresponding author upon reasonable request, subject to the original~licenses.}

\backmatteritem{Acknowledgments}{The
 authors thank the French National Center for Space Studies (CNES) for providing access to Pléiades imagery through PWH and for computational resources. We also acknowledge the IGN for providing LiDAR-HD data through their API.
The authors used Claude Sonnet 4.8, ChatGPT, and Gemini 3.1 Pro through their web interface for minor language correction and grammar editing during manuscript preparation. All content was reviewed and verified by the authors, who take full responsibility for the final text.}

\backmatteritem{Conflicts of Interest}{Authors Antoine Lorentz and Bastien Nespoulous were employed by the company Thales. The remaining authors declare that the research was conducted in the absence of any commercial or financial relationships that could be construed as a potential conflict of interest.}

\backmattersection{Abbreviations}{
The following abbreviations are used in this manuscript:\par\smallskip
\noindent
\begin{tabular}{@{}ll}
CN & ControlNet \\
DEM & Digital Elevation Model \\
DSM & Digital Surface Model \\
FFN & Feed-Forward Network \\
LiDAR & Light Detection and Ranging \\
MAE & Mean Absolute Error \\
MLP & Multilayer Perceptron \\
MM-DiT & Multimodal Diffusion Transformer \\
ODE & Ordinary Differential Equation \\
RGB & Red-Green-Blue \\
RMSE & Root Mean Square Error \\
SD3 & Stable Diffusion 3 \\
VAE & Variational Autoencoder
\end{tabular}
}

\FloatBarrier
\appendix
\titleformat{\section}{\large\bfseries}{Appendix \thesection.}{0.6em}{}
\titleformat{\subsection}{\normalsize\bfseries}{Appendix \thesubsection.}{0.6em}{}
\setcounter{table}{0}
\setcounter{figure}{0}
\renewcommand{\thetable}{A\arabic{table}}
\renewcommand{\thefigure}{A\arabic{figure}}

\section{Training and Inference Details}
\label{app:provenance}

Tables~\ref{tab:app_training} and~\ref{tab:app_fd_settings} summarize the training settings and the backbone evaluation procedure.

\begin{table}[!htbp]
\small
\caption{Common training hyperparameters used across the reported experiments.}
\label{tab:app_training}
\begin{tabularx}{\textwidth}{LL}
\toprule
\textbf{Hyperparameter} & \textbf{Value} \\
\midrule
Optimizer & AdamW \\
Learning-rate schedule & Constant \\
Weight decay & 0.01 \\
Mixed precision & BF16 \\
Gradient clipping & 1 \\
Timestep sampling & Uniform \\
Checkpoint interval & 1000 optimization steps \\
\bottomrule
\end{tabularx}
\end{table}

\begin{table}[!htbp]
\small
\caption{Settings for comparing pretrained and scratch backbones using FD$_{\mathrm{DINOv2}}$.}
\label{tab:app_fd_settings}
\begin{tabularx}{\textwidth}{Ll}
\toprule
\textbf{Setting} & \textbf{Value} \\
\midrule
Generated and reference samples & 50,000 each \\
Precision & BF16 \\
Sampling steps & 50 \\
Scheduler & FlowMatchEulerDiscreteScheduler \\
Scheduler shift & 1 \\
Classifier-free guidance & Disabled \\
Patch statistics & $s$ and $u$ from the evaluation LiDAR patch \\
Elevation preprocessing & Per-patch min--max normalization and channel replication \\
DINOv2 feature & \texttt{pooler\_output} \\
Pretrained learning rate & $1\times10^{-5}$ \\
Scratch learning-rate sweep & $\{1,2,5\}\times10^{-5}$; $\{1,2,4\}\times10^{-4}$ \\
Selected scratch learning rate & $2\times10^{-4}$ \\
\bottomrule
\end{tabularx}
\end{table}

\section{Computational Effect of Pruning}
\label{app:efficiency}

Table~\ref{tab:app_pruning_benchmark} compares the pruned and unpruned architectures, both with and without two ControlNets. Each test uses synthetic inputs and measures 50 denoising steps followed by VAE decoding into $512\times512$-pixel patches. The unpruned models receive synthetic text embeddings directly, without loading a text encoder. The timings exclude model loading, conditioning-image encoding, preprocessing, and storage. This isolates the computational effect of pruning rather than measuring the full processing pipeline or comparing prediction accuracy.

\begin{table}[!htbp]
\footnotesize
\caption{Runtime, throughput, and GPU memory for pruned and unpruned models on one NVIDIA A100 80\,GB PCIe GPU. Both use BF16, 50 steps with FlowMatchEulerDiscreteScheduler, and the same VAE decoder. Times are averages of two runs after one warm-up run. Memory is measured with \texttt{nvidia-smi} after warm-up.}

\label{tab:app_pruning_benchmark}
\setlength{\tabcolsep}{1.8pt}

\begin{adjustwidth}{-\extralength}{0cm}
\begin{tabularx}{\fulllength}{llRRRRRR}
\toprule
& & \multicolumn{3}{c}{\textbf{Pruned}} & \multicolumn{3}{c}{\textbf{Unpruned SD3 Medium}} \\
\textbf{Configuration} & \textbf{Batch} & \makecell{Time/Batch\\(s)} & \makecell{Throughput\\(Samples/s)} & \makecell{GPU Used\\(GiB)} & \makecell{Time/Batch\\(s)} & \makecell{Throughput\\(Samples/s)} & \makecell{GPU Used\\(GiB)} \\
\midrule
\multirow{4}{*}{Backbone + VAE} & 1  & 0.937 & 1.07 & 3.85 & 1.525 & 0.66 & 5.79 \\
 & 8  & 4.817 & 1.66 & 11.89 & 7.058 & 1.13 & 13.88 \\
 & 32 & 18.366 & 1.74 & 39.19 & 26.232 & 1.22 & 41.72 \\
 & 64 & 36.222 & 1.77 & 50.41 & 51.790 & 1.24 & 53.68 \\
\midrule
\multirow{4}{*}{Backbone + 2 ControlNets + VAE} & 1 & 1.874 & 0.53 & 6.33 & 3.089 & 0.32 & 10.25 \\
 & 8  & 10.219 & 0.78 & 15.46 & 14.749 & 0.54 & 19.51 \\
 & 32 & 38.871 & 0.82 & 46.51 & 54.877 & 0.58 & 50.75 \\
 & 64 & 76.751 & 0.83 & 63.05 & 108.343 & 0.59 & 67.46 \\
\bottomrule
\end{tabularx}
\end{adjustwidth}

\end{table}

Pruning reduces the transformer from 2.028B to 1.033B parameters and each ControlNet from 1.063B to 0.547B; the shared VAE contains 0.084B parameters. The pruned models are faster and use less memory at every tested batch size. Increasing the batch size from 32 to 64 brings little additional throughput while increasing memory use, making batch 32 a practical choice for this benchmark.

\section{Dataset Splits, Filtering, and Normalization}
\label{app:data_support}

Patches are extracted before assigning the training, validation, and test sets. Each city is divided into non-overlapping $512\times512$-pixel patches; incomplete patches at the image borders are dropped. Adjacent patches may touch, and no spatial buffer separates the splits. All RGB acquisitions covering the same patch remain in the same split. During training, one of these acquisitions is selected, so locations with more dates are not sampled more~often.

In each of the eight in-context cities, paired LiDAR--DSM--RGB patches are shuffled and divided into approximately 70\% training, 20\% test, and 10\% validation, with counts rounded to whole patches. All 3209 paired Bordeaux patches are reserved for testing. Table~\ref{tab:app_final_manifest_counts} gives the counts by city and training stage.

The RMSE filter removes DSM conditioning from the worst 10\% of eligible patches in each city. The LiDAR targets are retained when DSM conditioning is removed. This threshold, denoted P90, is based on error against LiDAR and is therefore target-dependent. Of the 40,589 patches assessed by this rule, 36,525 are retained and 4064 are excluded. The exclusions contain a larger share of Dense Urban and other Theia classes, and a smaller share of Sparse Built-Up pixels (Table~\ref{tab:p90_composition}). Thus, filtering changes the land-cover composition as well as input quality.

\begin{table}[H]
\footnotesize
\setlength{\tabcolsep}{3pt}
\caption{Numbers of geographic patches by city, available modalities, and split. $L$ denotes LiDAR, $D$ stereo DSM, and $R$ Pléiades RGB. The $L$ set is used for backbone training, $L+R$ for the RGB stage, and $L+D+R$ for paired training and evaluation. Multiple RGB dates at one location count as one patch. A zero indicates that no patches from that city are used in the corresponding set.}
\label{tab:app_final_manifest_counts}
\begin{tabularx}{\textwidth}{LRRRRRrr}
\toprule
& \multicolumn{1}{c}{\boldmath{$L$}} & \multicolumn{2}{c}{\boldmath{$L+R$}} & \multicolumn{4}{c}{\boldmath{$L+D+R$}} \\
\cmidrule{2-8}
\textbf{City} & \textbf{Train} & \textbf{Train }& \textbf{Val.} & \textbf{Train} & \textbf{Val.} & \makecell{\textbf{In-Context}\\\textbf{Test}} & \makecell{\textbf{Bordeaux}\\\textbf{Test}} \\
\midrule
Amiens & 24,591 & 17,350 & 2039 & 3555 & 507 & 1015 & 0 \\
Angers & 5380 & 0 & 0 & 0 & 0 & 0 & 0 \\
Arcachon & 9071 & 6127 & 713 & 1024 & 146 & 292 & 0 \\
Biarritz & 9599 & 5080 & 656 & 2927 & 417 & 835 & 0 \\
Bordeaux & 0 & 0 & 0 & 0 & 0 & 0 & 3209 \\
Clermont-Ferrand & 9311 & 0 & 0 & 0 & 0 & 0 & 0 \\
Grenoble & 8379 & 0 & 0 & 0 & 0 & 0 & 0 \\
Lyon & 10,104 & 0 & 0 & 0 & 0 & 0 & 0 \\
Marseille & 7288 & 0 & 0 & 0 & 0 & 0 & 0 \\
Montpellier & 47,648 & 11,160 & 1379 & 4392 & 627 & 1254 & 0 \\
Nancy & 9111 & 0 & 0 & 0 & 0 & 0 & 0 \\
Nantes & 10,556 & 0 & 0 & 0 & 0 & 0 & 0 \\
Nice & 16,801 & 10,483 & 1213 & 1540 & 220 & 440 & 0 \\
Paris & 42,356 & 17,738 & 2076 & 3328 & 475 & 950 & 0 \\
Poitiers & 7122 & 0 & 0 & 0 & 0 & 0 & 0 \\
Reims & 6102 & 0 & 0 & 0 & 0 & 0 & 0 \\
Rennes & 7872 & 0 & 0 & 0 & 0 & 0 & 0 \\
St-Etienne & 6174 & 0 & 0 & 0 & 0 & 0 & 0 \\
Strasbourg & 49,461 & 30,866 & 3507 & 2493 & 355 & 711 & 0 \\
Toulouse & 29,577 & 17,876 & 2084 & 3110 & 444 & 888 & 0 \\
\midrule
Total & 316,503 & 116,680 & 13,667 & 22,369 & 3191 & 6385 & 3209 \\
\bottomrule
\end{tabularx}
\end{table}

\begin{table}[H]
\small
\setlength{\tabcolsep}{4pt}
\caption{Land-cover composition before RMSE filtering and among excluded patches. Percentages are shares of pixels in each group. The ratio divides the excluded share by the original share: values above one indicate that a group is more common among exclusions. Other Theia classes includes all labels outside the five evaluation groups.}
\label{tab:p90_composition}
\begin{tabularx}{\textwidth}{LRRR}
\toprule
\textbf{Group} & \textbf{Before Filtering (\%)} & \textbf{Excluded Patches (\%)} & \textbf{Ratio (\boldmath{$\times$})} \\
\midrule
Dense urban & 5.56 & 10.03 & 1.81 \\
Croplands & 15.94 & 14.14 & 0.89 \\
Sparse built-up & 29.77 & 17.47 & 0.59 \\
Roads & 1.83 & 1.85 & 1.02 \\
Vegetation & 45.30 & 45.42 & 1.00 \\
Other Theia classes & 1.61 & 11.08 & 6.89 \\
\midrule
All pixels & 100.00 & 100.00 & 1.00 \\
\bottomrule
\end{tabularx}
\end{table}

\begin{figure}[H]
\centering
\includegraphics[width=0.86\textwidth]{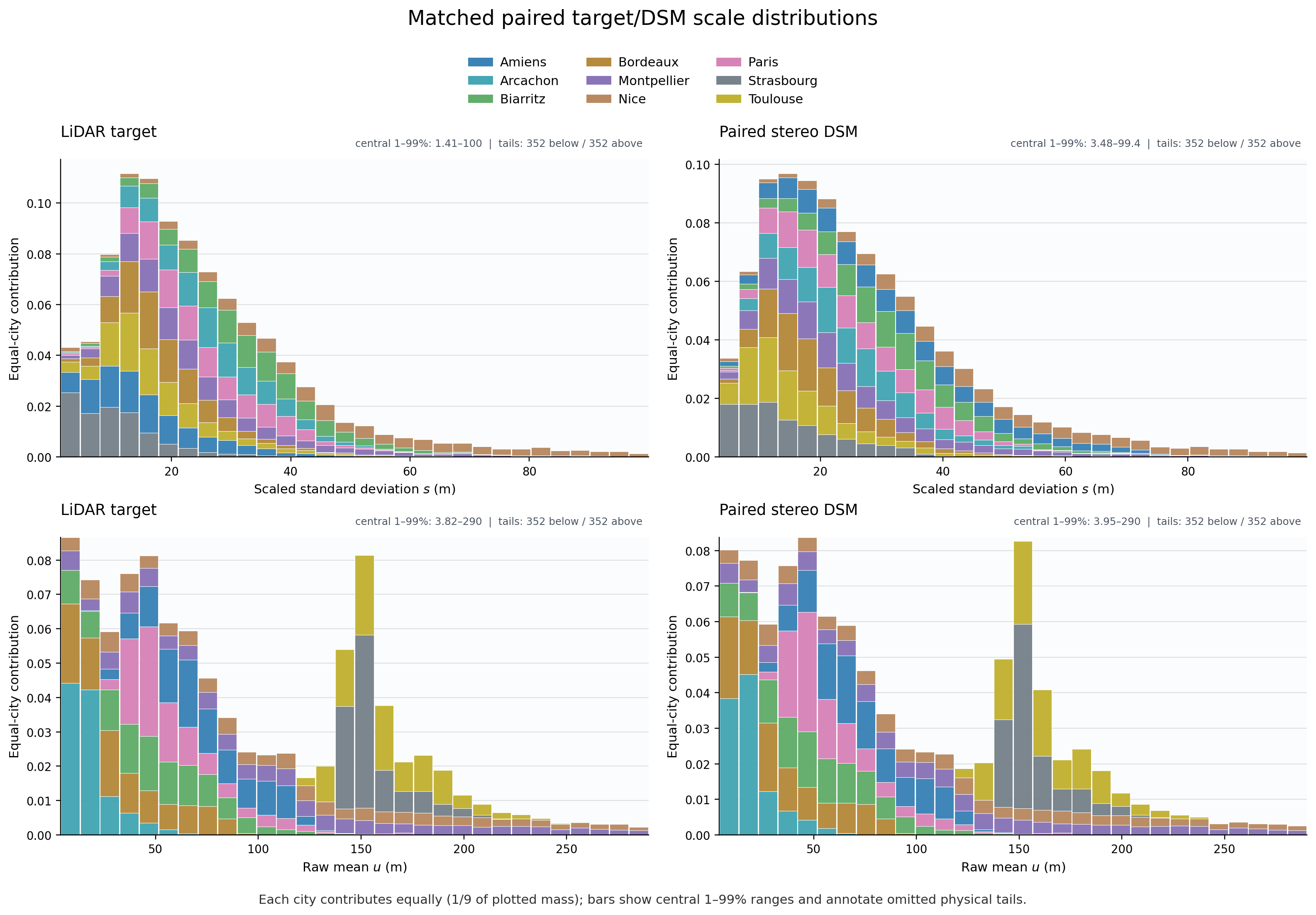}
\caption{Patch-wise normalization statistics for 35,154 LiDAR--DSM pairs from nine cities. The top row shows the scale $s=\operatorname{std}/0.25$; the bottom row shows the mean elevation $u$. LiDAR is shown on the left and the corresponding DSM on the right. Colors identify cities, weighted equally so that cities with more patches do not dominate the histograms. Each panel displays the central 98\% of values, with omitted extremes noted above the plot.}
\label{fig:app_scale_histograms}
\end{figure}

Figure~\ref{fig:app_scale_histograms} shows similar mean-elevation distributions for paired LiDAR and DSM patches, with differences in their local elevation variation. A broader scan of 329,288 LiDAR patches and 35,154 paired DSM patches found no zero standard deviations. The smallest LiDAR standard deviation was 0.00473\,m in a nearly flat water patch; 439 LiDAR patches had standard deviations below 0.1\,m, mostly over water. No paired DSM patch fell below this threshold. These observations describe the dataset; the implementation does not clip the scale or provide a zero-variance fallback.

Land-cover evaluation uses five groups from Theia Land Cover 2021: Dense Urban, Croplands, Sparse Built-Up, Roads, and Vegetation. The 10\,m labels are resampled to 0.5\,m by nearest-neighbor sampling, so each source label covers approximately $20\times20$ evaluation pixels. Classes outside these five groups are not included in the reported model comparisons.

\FloatBarrier
\section{Additional Evaluation Results}
\label{app:evaluation}

\subsection{Performance on Patches Excluded by the RMSE Filter}
Of the 4064 patches excluded by the P90 filter, 4046 have Pléiades imagery and can be evaluated with both conditioning configurations. We compare the calibrated DSM, DSM-only, and DSM~+~RGB on the same pixels, using only locations with valid elevations in both the input DSM and LiDAR. Predictions use the standard inference procedure with 20~seeds (0--19).

\begin{table}[!htbp]
\footnotesize
\setlength{\tabcolsep}{3pt}
\caption{RMSE on patches excluded by the P90 filter, calculated over all valid LiDAR/DSM pixels in each land-cover group. Values are in meters; model results are mean~$\pm$~standard deviation over 20~inference seeds.}
\label{tab:stress_landcover}

\begin{adjustwidth}{-\extralength}{0cm}
\begin{tabularx}{\fulllength}{LCCCCC}
\toprule
& \multicolumn{5}{c}{\textbf{Land Cover}} \\
\cmidrule(lr){2-6}
\textbf{Method} & \textbf{Dense Urban} & \textbf{Croplands} & \textbf{Sparse Built-Up} & \textbf{Roads} & \textbf{Vegetation} \\
\midrule
\multicolumn{6}{l}{\textbf{Patches excluded by the RMSE filter}} \\
Calibrated stereo DSM (CARS) & 9.686 & 23.050 & 5.564 & 5.740 & 9.804 \\
\midrule
Ours---DSM only & 6.179 $\pm$ 0.028 & 2.064 $\pm$ 0.019 & 4.776 $\pm$ 0.010 & 4.162 $\pm$ 0.015 & 6.849 $\pm$ 0.023 \\
Ours---DSM + RGB & 5.736 $\pm$ 0.032 & 2.240 $\pm$ 0.021 & 4.464 $\pm$ 0.017 & 3.743 $\pm$ 0.028 & 6.718 $\pm$ 0.035 \\
\bottomrule
\end{tabularx}
\end{adjustwidth}

\end{table}

Both models reduce RMSE relative to the calibrated input in every group. RGB brings additional gains except in Croplands, where DSM-only performs better. These results extend the evaluation to high-error inputs on pixels with valid DSM elevations.

\subsection{Recovery of Original DSM Voids}
\label{app:void_recovery}

We evaluate void recovery on the retained in-context and Bordeaux test sets. Only pixels with a valid LiDAR elevation but no input DSM elevation are included. Missing pixels outside the source DSM image are excluded, so image borders are not treated as voids. The calibrated DSM cannot serve as a numerical baseline here because it has no elevation at these locations.

\begin{table}[!htbp]
\footnotesize
\setlength{\tabcolsep}{3pt}
\caption{RMSE within original DSM voids on the retained test sets. Errors are calculated over all evaluated void pixels in each land-cover group, excluding areas outside the source DSM image. Values are mean $\pm$ standard deviation in meters over 20 inference seeds.}
\label{tab:void_landcover}

\begin{adjustwidth}{-\extralength}{0cm}
\begin{tabularx}{\fulllength}{LCCCCC}
\toprule
& \multicolumn{5}{c}{\textbf{Land Cover}} \\
\cmidrule(lr){2-6}
\textbf{Method} & \textbf{Dense Urban} & \textbf{Croplands} & \textbf{Sparse Built-Up} & \textbf{Roads} & \textbf{Vegetation} \\
\specialrule{0.08em}{0.4em}{0.3em}
\multicolumn{6}{l}{\textbf{In-context}} \\
\addlinespace[0.3em]
Ours---DSM only & 4.135 $\pm$ 0.038 & 1.179 $\pm$ 0.050 & 4.747 $\pm$ 0.056 & 3.720 $\pm$ 0.096 & 6.040 $\pm$ 0.029 \\
Ours---DSM + RGB & 3.576 $\pm$ 0.037 & 0.932 $\pm$ 0.023 & 3.801 $\pm$ 0.047 & 3.215 $\pm$ 0.083 & 5.362 $\pm$ 0.030 \\
\specialrule{0.08em}{0.4em}{0.3em}
\multicolumn{6}{l}{\textbf{Held-out Bordeaux}} \\
\addlinespace[0.3em]
Ours---DSM only & 4.147 $\pm$ 0.083 & 3.856 $\pm$ 0.290 & 4.995 $\pm$ 0.029 & 3.309 $\pm$ 0.085 & 5.916 $\pm$ 0.030 \\
Ours---DSM + RGB & 3.901 $\pm$ 0.103 & 3.402 $\pm$ 0.060 & 4.385 $\pm$ 0.031 & 2.944 $\pm$ 0.164 & 5.209 $\pm$ 0.037 \\
\bottomrule
\end{tabularx}
\end{adjustwidth}

\end{table}

DSM~+~RGB has lower mean void RMSE than DSM-only in every group in both test sets. Vegetation remains among the most difficult groups for both models.

\subsection{Spatial Uncertainty and Inference Seeds}
The confidence intervals in Figure~\ref{fig:landcover_rmse_ci} are based on 10,000 paired bootstrap resamples. We resample the eight in-context cities, or $2\times2$-patch blocks within Bordeaux, using the same sampled areas for all methods. In each resample, we calculate RMSE separately for each inference seed, average the 20 values for each model, and then calculate the differences between methods. This follows the same averaging procedure as Table~\ref{tab:evaluation}. The intervals describe spatial variation while keeping the inference seeds fixed.

Table~\ref{tab:seed_sd} separately reports the variation of MAE and RMSE across inference seeds; the largest standard deviation is 0.035\,m. This includes random generation and, for DSM~+~RGB, selection among available RGB acquisitions. Neither analysis includes repeated training~runs.

\begin{table}[!htbp]
\footnotesize
\setlength{\tabcolsep}{3pt}
\caption{Standard deviations of the MAE and RMSE reported in Table~\ref{tab:evaluation}, measured across 20~inference seeds. Evaluation uses pixels with valid LiDAR and input DSM elevations. All values are in~meters.}
\label{tab:seed_sd}
\begin{adjustwidth}{-\extralength}{0cm}
\begin{tabularx}{\fulllength}{lCCCCCCCCCC}
\toprule
& \multicolumn{10}{c}{\textbf{Land Cover}} \\
\cmidrule{2-11}
\textbf{Method}
& \multicolumn{2}{c}{\textbf{Dense Urban}}
& \multicolumn{2}{c}{\textbf{Croplands}}
& \multicolumn{2}{c}{\textbf{Sparse Built-Up}}
& \multicolumn{2}{c}{\textbf{Roads}}
& \multicolumn{2}{c}{\textbf{Vegetation}} \\
\cmidrule{2-11}

& \makecell{\textbf{MAE}\\\textbf{SD}} & \makecell{\textbf{RMSE}\\\textbf{SD}}
& \makecell{\textbf{MAE}\\\textbf{SD}} & \makecell{\textbf{RMSE}\\\textbf{SD}}
& \makecell{\textbf{MAE}\\\textbf{SD}} & \makecell{\textbf{RMSE}\\\textbf{SD}}
& \makecell{\textbf{MAE}\\\textbf{SD}} & \makecell{\textbf{RMSE}\\\textbf{SD}}
& \makecell{\textbf{MAE}\\\textbf{SD}} & \makecell{\textbf{RMSE}\\\textbf{SD}} \\
\midrule
\multicolumn{11}{l}{\textbf{In-context}} \\
Ours---DSM only
& 0.005 & 0.006
& 0.003 & 0.007
& 0.003 & 0.005
& 0.011 & 0.024
& 0.002 & 0.004 \\
Ours---DSM + RGB
& 0.007 & 0.016
& 0.003 & 0.005
& 0.004 & 0.006
& 0.008 & 0.017
& 0.003 & 0.007 \\
\midrule
\multicolumn{11}{l}{\textbf{Held-out Bordeaux}} \\
Ours---DSM only
& 0.011 & 0.024
& 0.025 & 0.030
& 0.003 & 0.005
& 0.014 & 0.027
& 0.005 & 0.009 \\
Ours---DSM + RGB
& 0.012 & 0.025
& 0.020 & 0.024
& 0.004 & 0.007
& 0.015 & 0.035
& 0.006 & 0.010 \\
\bottomrule
\end{tabularx}
\end{adjustwidth}
\end{table}

\subsection{Bias and Error Distribution}
\begin{figure}[H]
\centering

\includegraphics[width=0.82\textwidth]{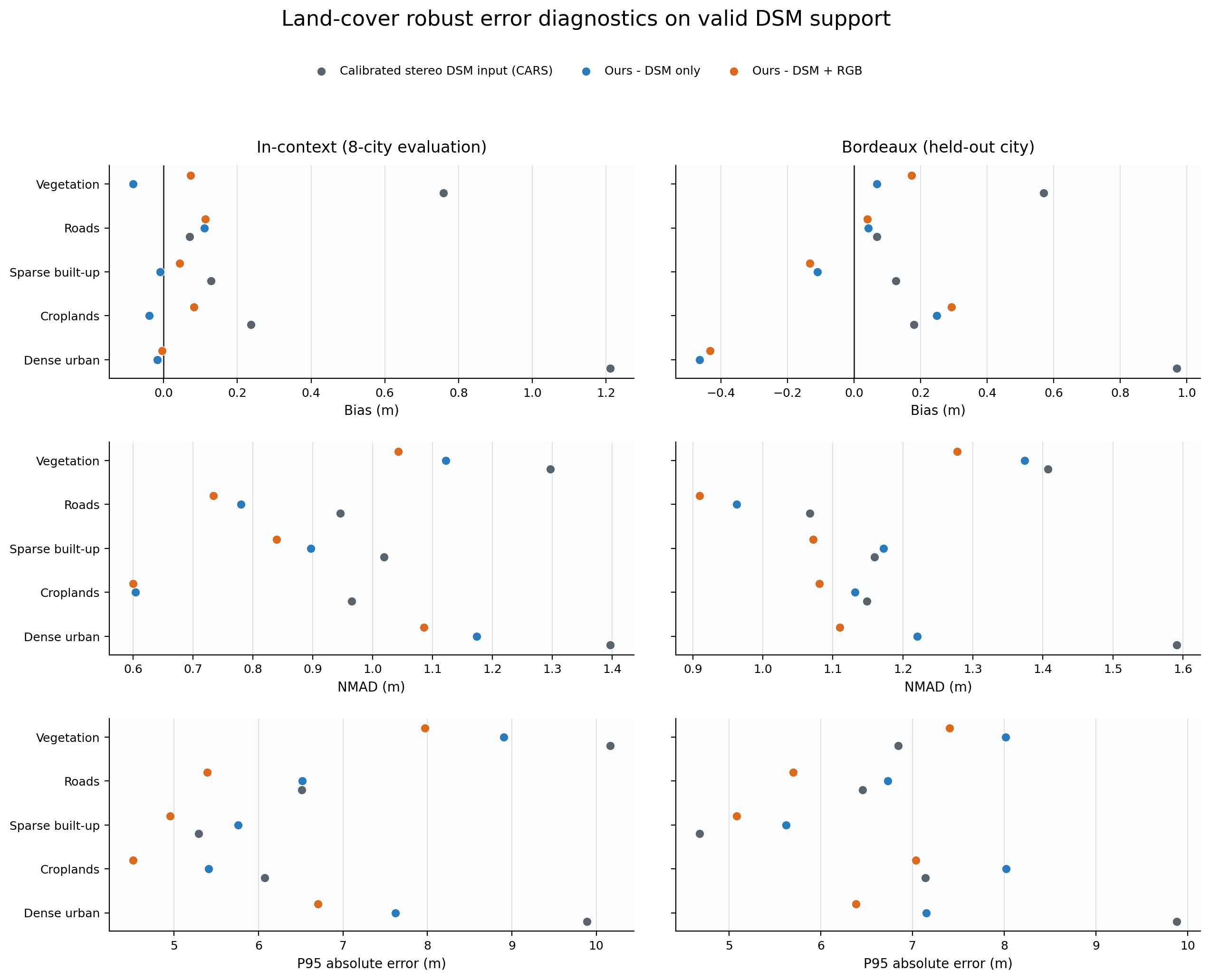}
\caption{Signed bias, NMAD, and P95 absolute error by land cover for the in-context test (\textbf{left}) and Bordeaux (\textbf{right}). Only pixels with valid LiDAR and input DSM elevations are evaluated. Bias uses all evaluated pixels; NMAD and P95 are estimated by sampling 256 pixels per patch, giving each patch equal weight before grouping by land cover. All values are in meters.}
\label{fig:app_robust_errors}
\end{figure}

Figure~\ref{fig:app_robust_errors} complements MAE and RMSE with three measures. Signed bias indicates whether elevations are overestimated or underestimated. The normalized median absolute deviation (NMAD) measures the spread of errors around their median with less sensitivity to outliers, while P95 is the 95th percentile of absolute error. DSM~+~RGB lowers NMAD in every group in both tests. Its P95 is also lower than the calibrated input in all in-context groups, but remains higher for Sparse Built-Up and Vegetation in Bordeaux.

\subsection{Effect of Excluding Land-Cover Boundaries}
\label{app:sensitivity}

\begin{figure}[H]
\centering
\includegraphics[width=0.82\textwidth]{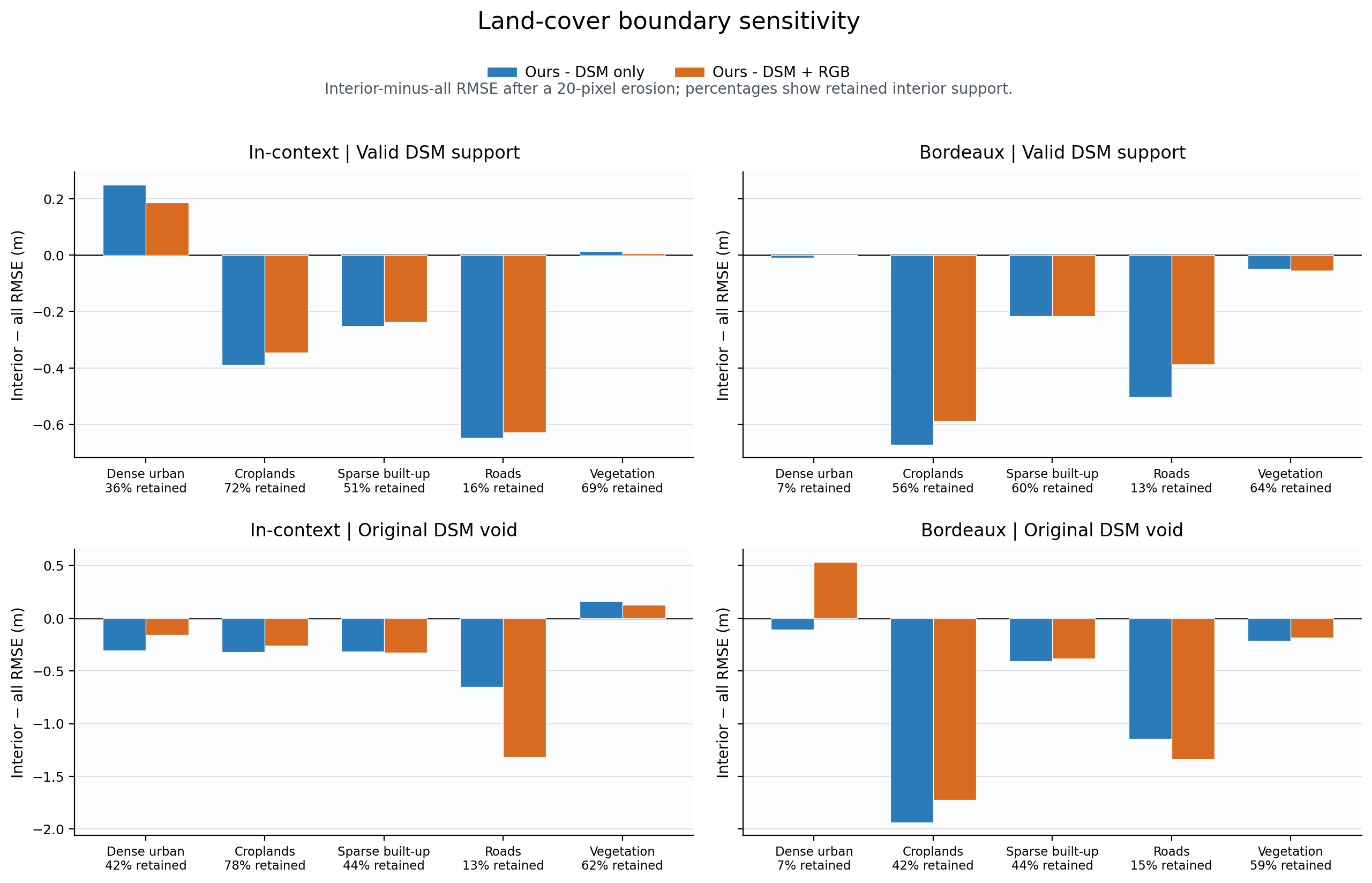}
\caption{Change in RMSE after excluding land-cover boundaries and patch edges. Bars show RMSE for the remaining interior pixels minus RMSE for all evaluated pixels; negative values mean lower interior error. The top row evaluates valid DSM pixels and the bottom row evaluates original DSM voids within the source image. Percentages indicate the fraction of pixels retained in each group. Errors are in meters.}
\label{fig:app_boundary_sensitivity}
\end{figure}

To assess the effect of mixed labels near land-cover boundaries, we repeat the evaluation after removing pixels within a 20-pixel (10\,m) square buffer of class boundaries and patch edges. Figure~\ref{fig:app_boundary_sensitivity} compares RMSE before and after this removal, separately for valid DSM pixels and original voids.

Removing boundaries generally lowers RMSE for Croplands, Sparse Built-Up, and Roads. The effect is less uniform for Dense Urban and Vegetation. The percentages below the bars show how much of each group remains: only a small fraction of Roads and Bordeaux Dense Urban pixels survives this exclusion. The analysis therefore describes class interiors, not improved label resolution.

\FloatBarrier
\section{Selected Qualitative Cases}
\label{app:qualitative}

\begin{figure}[!htbp]
\centering
\resizebox{0.80\textwidth}{!}{
\begin{tikzpicture}
    \node[anchor=south west] (img) at (0,0)
        {\includegraphics[width=\linewidth]{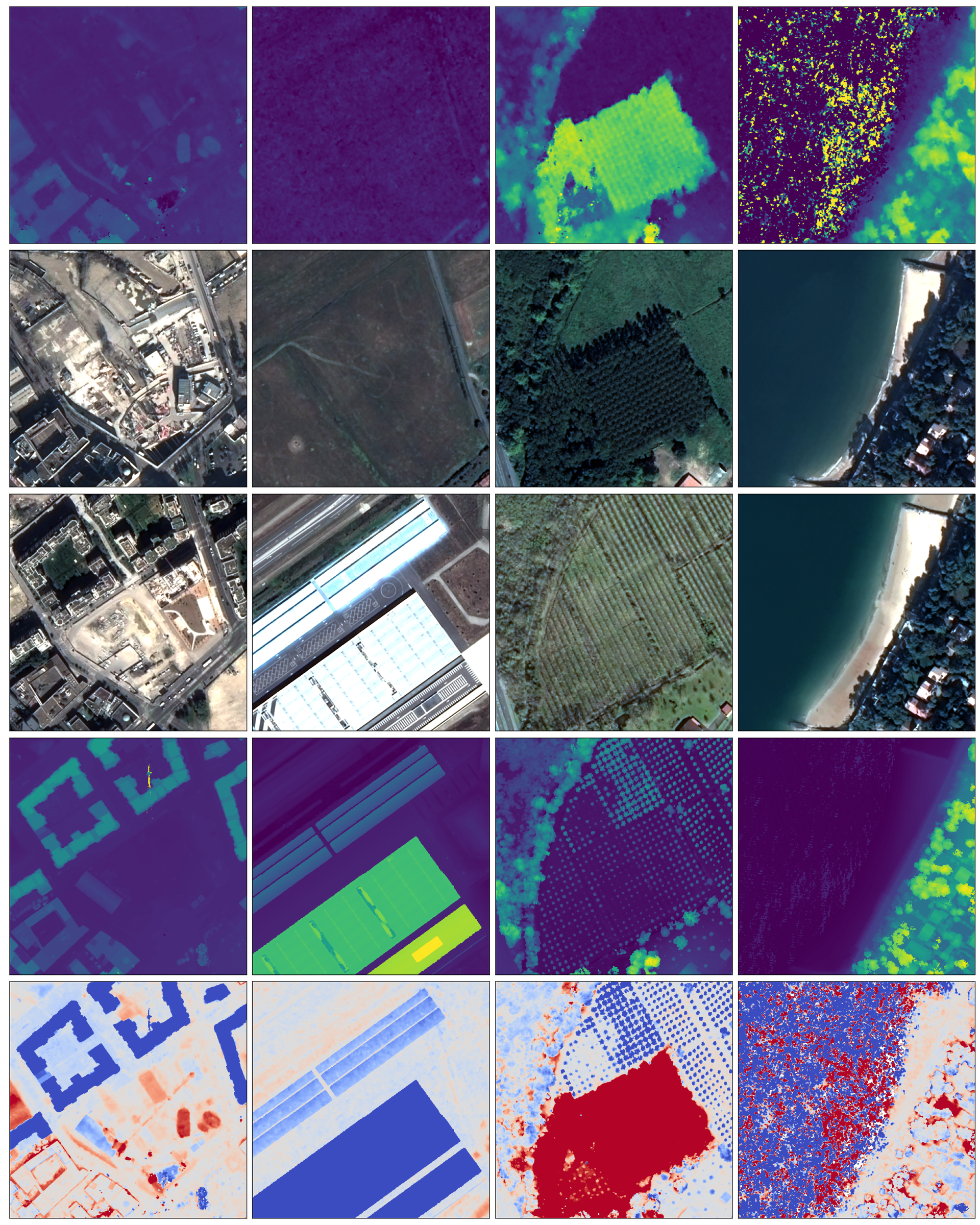}};
    \begin{scope}[x={(img.south east)}, y={(img.north west)}]
        \node[anchor=center, align=center] at (0.125,1.02) {Redevelopment\\Paris};
        \node[anchor=center, align=center] at (0.375,1.02) {Industrial\\development\\Toulouse};
        \node[anchor=center, align=center] at (0.625,1.02) {Woodland-to-orchard\\Biarritz};
        \node[anchor=center, align=center] at (0.875,1.02) {Water/beach state\\Arcachon};
        \node[anchor=center, rotate=90] at (-0.05,0.9) {DSM (CARS)};
        \node[anchor=center, rotate=90, align=center] at (-0.05,0.7) {Selected\\Pléiades};
        \node[anchor=center, rotate=90, align=center] at (-0.05,0.5) {Later\\Pléiades};
        \node[anchor=center, rotate=90] at (-0.05,0.3) {LiDAR};
        \node[anchor=center, rotate=90, align=center] at (-0.05,0.1) {Error map\\DSM};
    \end{scope}
\end{tikzpicture}
}
\vspace{2pt}
\begin{tikzpicture}
    \node[anchor=east] at (0, 0) {\small Legend (m)};
    \node[anchor=west] (cbar) at (0, 0)
        {\includegraphics[width=0.8\linewidth]{figs/coolwarm.png}};
    \foreach \val/\pos in {-10/0, -5/0.25, 0/0.5, 5/0.75, 10/1.0}{
        \node[anchor=north] at ($(cbar.south west)!\pos!(cbar.south east)$)
            {\small $\val$};
    }
\end{tikzpicture}
\caption{Excluded patches with visible changes between RGB dates: Paris (2019/2025), Toulouse (2012/2022), Biarritz (2014/2025), and Arcachon (2021/2024). Rows show the calibrated DSM, conditioning RGB, later RGB, LiDAR, and DSM error. Within each column, DSM and LiDAR share the LiDAR elevation range; errors use a fixed $-10$ to $+10$\,m scale. Pléiades ©CNES 2012/2014/2019/2021/2022/2024/2025, Distribution CNES PWH; LiDAR-HD ©IGN.}
\label{fig:app_original_stress}
\end{figure}

Figure~\ref{fig:app_original_stress} compares RGB acquisitions at four locations excluded by the RMSE filter. They cover urban redevelopment, industrial construction, woodland converted to an orchard, and a changing shoreline. The RGB comparisons illustrate changes in surface appearance.

\begin{figure}[p]
\centering

\resizebox{0.94\textwidth}{!}{
\begin{tikzpicture}
    \node[anchor=south west] (img) at (0,0) {\includegraphics[width=\linewidth]{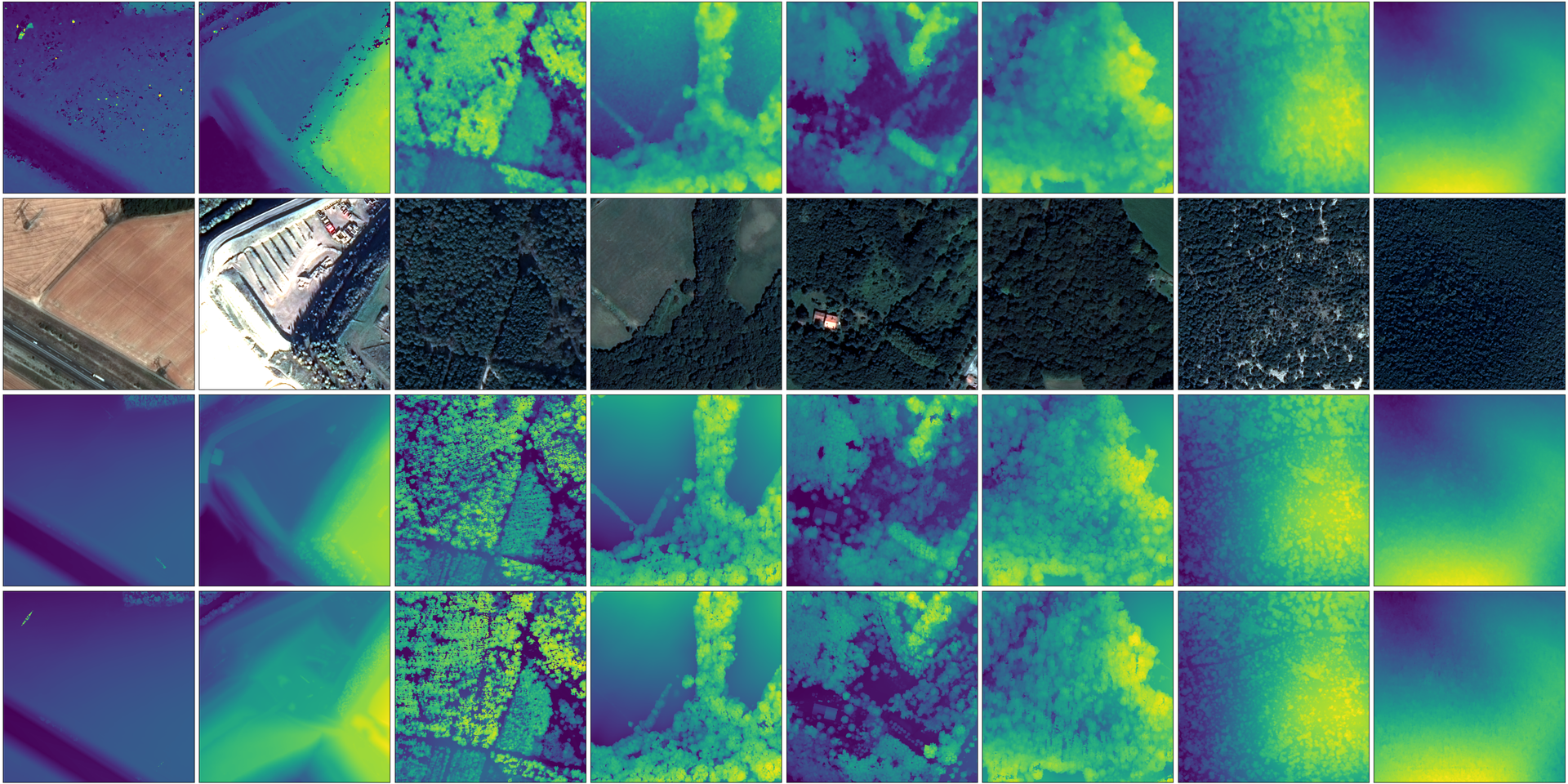}};
    \begin{scope}[x={(img.south east)}, y={(img.north west)}]
        \node[anchor=east, font=\small] at (-0.012,0.875) {DSM};
        \node[anchor=east, font=\small] at (-0.012,0.625) {RGB};
        \node[anchor=east, font=\small] at (-0.012,0.375) {DSM~+~RGB};
        \node[anchor=east, font=\small] at (-0.012,0.125) {LiDAR};
    \end{scope}
\end{tikzpicture}}
\vspace{2pt}\hrule\vspace{2pt}
\resizebox{0.90\textwidth}{!}{
\begin{tikzpicture}
    \node[anchor=south west] (img) at (0,0) {\includegraphics[width=\linewidth]{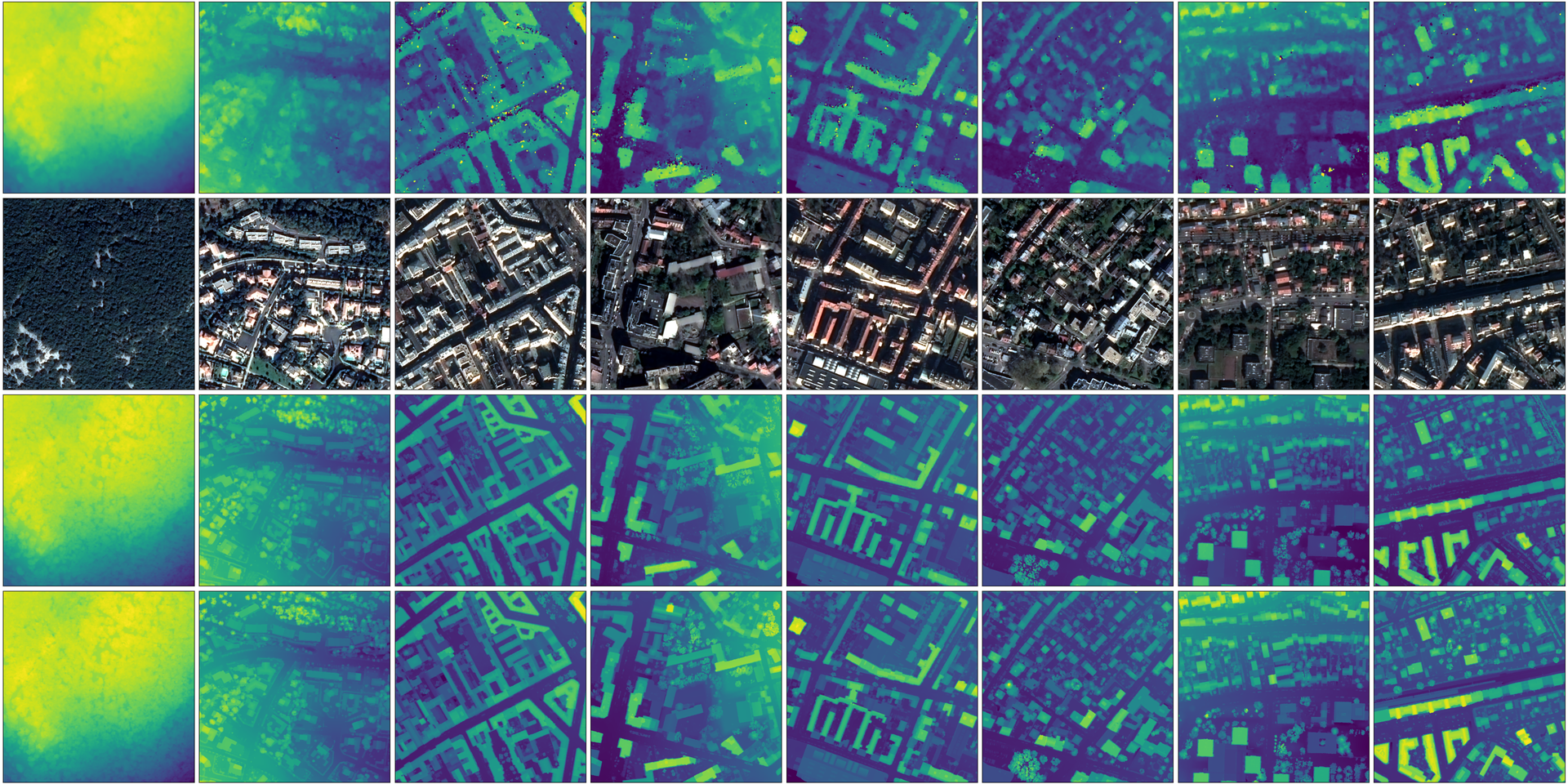}};
    \begin{scope}[x={(img.south east)}, y={(img.north west)}]
        \node[anchor=east, font=\small] at (-0.012,0.875) {DSM};
        \node[anchor=east, font=\small] at (-0.012,0.625) {RGB};
        \node[anchor=east, font=\small] at (-0.012,0.375) {DSM~+~RGB};
        \node[anchor=east, font=\small] at (-0.012,0.125) {LiDAR};
    \end{scope}
\end{tikzpicture}}
\vspace{2pt}\hrule\vspace{2pt}
\resizebox{0.90\textwidth}{!}{
\begin{tikzpicture}
    \node[anchor=south west] (img) at (0,0) {\includegraphics[width=\linewidth]{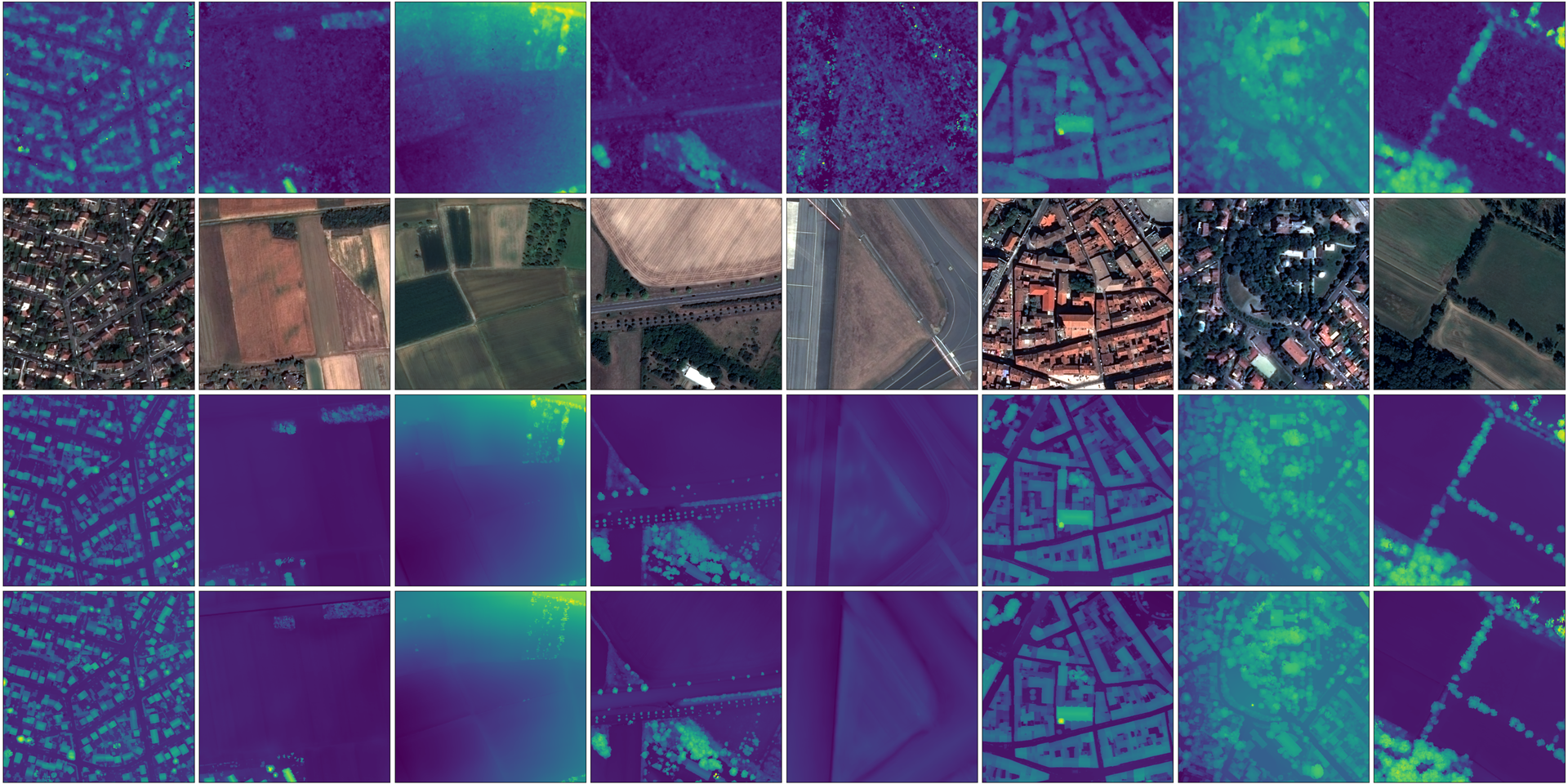}};
    \begin{scope}[x={(img.south east)}, y={(img.north west)}]
        \node[anchor=east, font=\small] at (-0.012,0.875) {DSM};
        \node[anchor=east, font=\small] at (-0.012,0.625) {RGB};
        \node[anchor=east, font=\small] at (-0.012,0.375) {DSM~+~RGB};
        \node[anchor=east, font=\small] at (-0.012,0.125) {LiDAR};
    \end{scope}
\end{tikzpicture}}
\caption{Twenty-four patches drawn at random without replacement from the 6385 in-context test patches, without visual or error-based selection. Each panel contains eight examples, with rows showing the calibrated DSM, RGB, DSM~+~RGB prediction, and LiDAR. Within each column, all elevation maps share the LiDAR range. Pléiades ©CNES 2012--2024, Distribution CNES PWH; LiDAR-HD ©IGN.}
\label{fig:app_random_retained}
\end{figure}

\begin{figure}[p]
\centering
\resizebox{0.85\textwidth}{!}{
\begin{tikzpicture}
    \node[anchor=south west] (img) at (0,0)
        {\includegraphics[width=\linewidth]{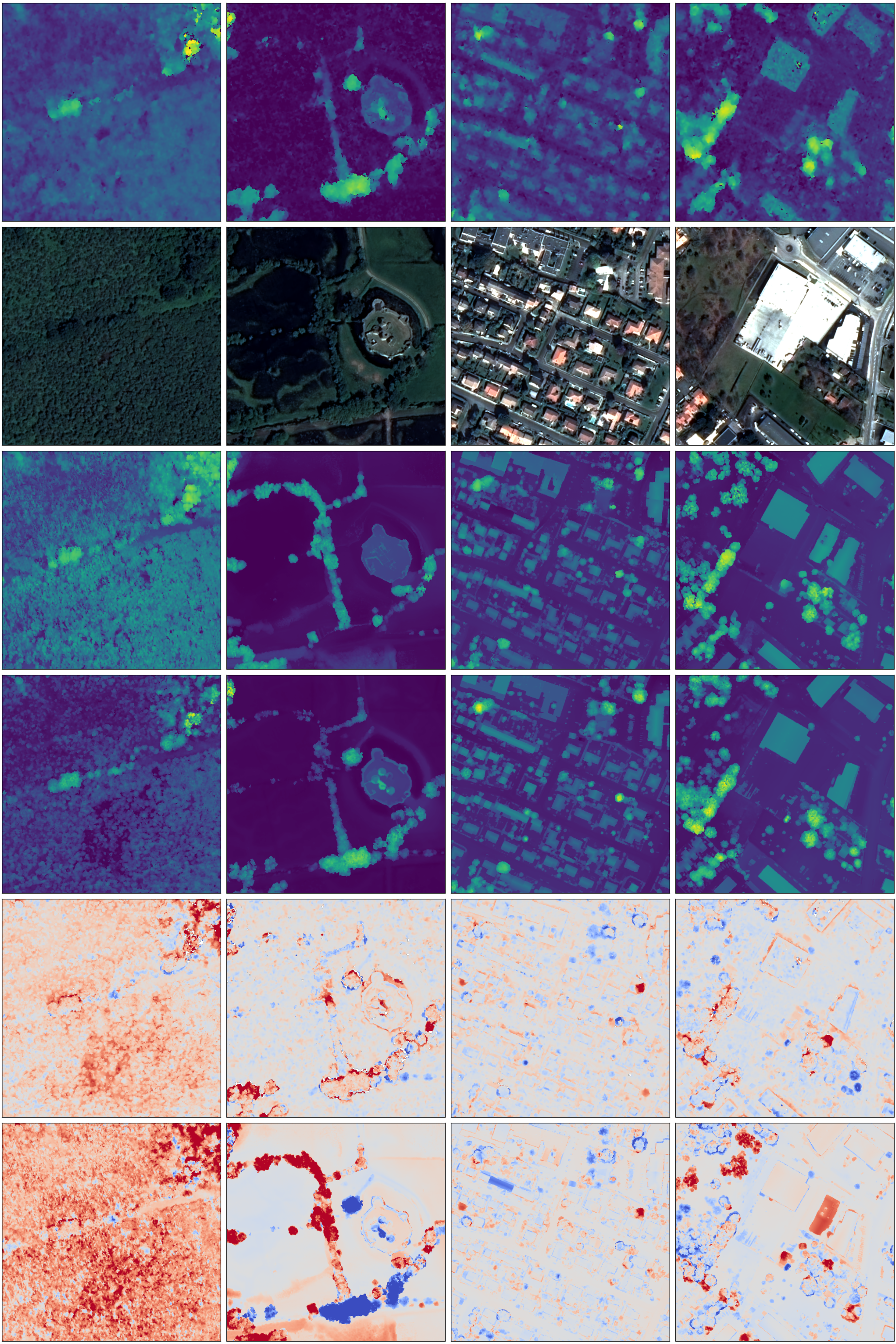}};
    \begin{scope}[x={(img.south east)}, y={(img.north west)}]
        \node[anchor=center, align=center] at (0.125,1.02) {Whole-patch\\case 1};
        \node[anchor=center, align=center] at (0.375,1.02) {Whole-patch\\case 2};
        \node[anchor=center, align=center] at (0.625,1.02) {Dense Urban\\case 1};
        \node[anchor=center, align=center] at (0.875,1.02) {Dense Urban\\case 2};
        \node[anchor=center, rotate=90] at (-0.05,0.917) {DSM (CARS)};
        \node[anchor=center, rotate=90] at (-0.05,0.750) {Selected Pléiades};
        \node[anchor=center, rotate=90, align=center] at (-0.05,0.583) {DSM + Pléiades\\ControlNet};
        \node[anchor=center, rotate=90] at (-0.05,0.417) {LiDAR};
        \node[anchor=center, rotate=90, align=center] at (-0.05,0.250) {Error map\\DSM};
        \node[anchor=center, rotate=90, align=center] at (-0.05,0.083) {Error map\\DSM + Pléiades};
    \end{scope}
\end{tikzpicture}
}
\vspace{2pt}
\begin{tikzpicture}
    \node[anchor=east] at (0, 0) {\small Legend (m)};
    \node[anchor=west] (cbar) at (0, 0)
        {\includegraphics[width=0.8\linewidth]{figs/coolwarm.png}};
    \foreach \val/\pos in {-10/0, -5/0.25, 0/0.5, 5/0.75, 10/1.0}{
        \node[anchor=north] at ($(cbar.south west)!\pos!(cbar.south east)$)
            {\small $\val$};
    }
\end{tikzpicture}
\caption{Bordeaux examples where DSM~+~RGB increases RMSE relative to the calibrated input. The first two columns show the largest whole-patch increases; the last two show the largest increases within Dense Urban pixels. Selection uses seed-0 predictions and pixels with valid LiDAR and DSM elevations. Rows show the DSM, RGB, prediction, LiDAR, DSM error, and prediction error. Within each column, elevation maps share the LiDAR range; both error rows use a fixed $-10$ to $+10$\,m scale. These are selected failure cases, not average results. Pléiades ©CNES 2016/2020/2025, Distribution CNES PWH; LiDAR-HD ©IGN.}
\label{fig:app_negative_transfer}
\end{figure}

The random examples in Figure~\ref{fig:app_random_retained} show smoother open surfaces and more distinct buildings and tree crowns, alongside remaining differences from LiDAR. Figure~\ref{fig:app_negative_transfer} documents local errors in vegetation and built areas despite the aggregate improvements. These visual comparisons complement the land-cover metrics; the broader limitations are discussed in Section~\ref{sec:limitations}.

\FloatBarrier
\section{Access and Limitations}
\label{app:access}

IGN/CNES and commercial-imagery licenses restrict redistribution of the data. The paper provides rendered examples and numerical summaries, but no source imagery, elevation rasters, or model weights. Access to the underlying products remains subject to the original licenses, and the implementation is not publicly released. The results apply to vertically co-registered, compatible DSM--LiDAR pairs in the evaluated French cities, as discussed in Section~\ref{sec:limitations}.

\begin{adjustwidth}{-\extralength}{0cm}

\end{adjustwidth}

\end{document}